\documentstyle{article}

\title{Heat equilibrium  distribution in a  turbulent
flow}
\author{ Z. Haba\\Institute of Theoretical Physics, University of Wroclaw,
\\50-204 Wroclaw, Plac Maxa Borna 9,Poland\\e-mail:zhab@ift.uni.wroc.pl}
\date{PACS:47.10.A-,44.27.+g,47.27.eb}
\begin{document}
\maketitle
\begin{abstract}
We consider a shear flow of a  scale invariant  Gaussian random
velocity field which does not depend on the coordinates in the
direction of the flow. We investigate a heat advection coming from
a Gaussian random homogeneous source. We discuss a relaxation at
large time of a temperature distribution determined by the forced
advection-diffusion equation. We represent the temperature
correlation functions by means of the Feynman-Kac formula. Jensen
inequalities are applied for lower and upper bounds on the
correlation functions.
 We show
that at finite time there is no velocity dependence of long range
temperature correlations (low momentum asymptotics) in the
direction of the flow but the equilibrium heat distribution has
large distance correlations (low momentum behavior) with an index
depending on the scaling index of the random flow and of the index
of the random forcing. If the velocity has correlations growing
with the distance ( a turbulent flow) then the large distance
correlations depend in a crucial way on the scaling index of the
turbulent flow. In such a case the correlations increase in the
direction of the flow and decrease in the direction perpendicular
to the flow making the stream of heat more coherent.

\end{abstract}
\section{Introduction}
We investigate a heat advection in a random flow which is supposed
to be "turbulent".  The  turbulence is a complex phenomenon
difficult to define and avoiding a description in precise
mathematical terms. The complexity of turbulence can be related to
its dependence on the length scale relevant for undergoing
experiments. In this paper  we apply only some aspects of the
turbulent flow: randomness of the velocity field, its
self-similarity and long range correlations . The appearance of
the turbulence should have an impact on transport phenomena
described by an advection-diffusion equation of a passive scalar
\cite{nature}. Such an equation can describe a transport of heat,
a mass or some impurities. We are interested in the equilibrium
distribution of solutions of the random advection-diffusion
equation . The equilibrium is possible only under an external
forcing (a heat source). We are interested in the equilibrium
distribution at all scales. Such an equilibrium will depend on the
forcing. The universality is possible only in the inertial range
\cite{kolmogorov}\cite{frisch} \cite{gawedzki} where the external
forcing should not be relevant(see ref.\cite{staicu} for some
recent shear flow experiments) . Although the precise equilibrium
distribution depends on the form of the forcing the asymptotic
behavior of correlation functions depends solely on the asymptotic
behavior of the random forcing. We investigate the way the long
range correlations of the fluid velocity influence the long range
correlations of the temperature.

We assume that there is a distinguished direction of the fluid
velocity ${\bf V}$. We make a decomposition $X=({\bf x},{\bf
z})\in R^{D}$ with ${\bf x}\in R^{d}$ and ${\bf z}\in R^{D-d}$ ;
${\bf V}(\tau,{\bf x})$ depends only on ${\bf x}\in R^{d}$ and has
the non-vanishing components only in $R^{D-d}$ (in such a case it
satisfies automatically $\nabla {\bf V}=0$;for physical
applications $D=3$ and $d=2$ or $d=1$). As a typical example we
could consider a fluid flow  $V_{z}(x,y)$ in the direction of the
$z$-axis which does not depend on $z$. We can impose such an
anisotropy of the flow by an external force ${\bf R}$ which
depends only on ${\bf x}$ and has non-zero components solely in
the ${\bf z}$ direction. So, we consider the Navier-Stokes
equation with such a random force ${\bf R}$

\begin{displaymath}
\partial_{t}{\bf V}+{\bf V}\nabla {\bf V}-\nu \triangle {\bf V}
={\bf R}
\end{displaymath}
The $({\bf 0},{\bf V}({\bf x}))$ solution of the Navier-Stokes
equation is the solution of the linear equation ( for the ${\bf
z}$-component)
\begin{displaymath}
\partial_{t}{\bf V}-\nu \triangle_{\bf x} {\bf V}
={\bf R}
\end{displaymath}( together with a zero solution for the ${\bf
x}$-component). By a proper  choice of the external force ${\bf
R}$ we can simulate a large class of ${\bf x}$-dependent flows.

In secs.2-3 we discuss the advection-diffusion equation, the
random velocity and a random forcing.
  The advection-diffusion equation can be
solved by means of the Feynman-Kac formula. The Feynman-Kac
solution  has already been discussed by other authors
\cite{majda}-\cite{glimm}. These authors have been interested  in
the asymptotic behavior of the advection-diffusion equation
without forcing. Our main interest (secs.4-5) is in the asymptotic
behavior for large time and distances of correlation functions of
the temperature field resulting from the advection-diffusion
equation with forcing describing the heat injection. First, in
sec.3 we simulate forcing by a constant gradient term in the
temperature. We obtain a simple soluble model of advection
illustrating some general features. In general, we can obtain some
lower and upper bounds on the correlation functions by means of
the Jensen inequalities (sec.5). For the sake of  simplicity we
concentrate on the two-point correlations. In sec.6 we show how
our methods can be extended to multi-point correlations. We obtain
 asymptotic behavior of the Fourier transform of the correlation
functions for small and large momenta.
 We compare our methods
and results (in secs.4-6 and in the Appendix B)  with  an exactly
soluble model of Kraichnan
\cite{kraichnan}\cite{gawedzki}\cite{falk} (defined by a velocity
field which is a white noise in time). The random advection is
closely connected with a diffusion. In fact, under some natural
assumptions random advection enforces diffusion
\cite{kesten}\cite{ave}\cite{komorowski} and vice versa the
diffusion can be expressed as a white noise advection \cite{verg}.
However, when we choose no diffusion (zero molecular diffusivity)
 in the initial equation of advection describing the temperature evolution then we
 obtain a
model of advection (discussed in Appendix A) as a limit of the
solution of the random advection-diffusion equation. The limit of
zero molecular diffusivity has been discussed earlier in
refs.\cite{wei}\cite{hula}.

 In the text some positive constants arise
(denoted usually as $K$,$c_{1}$, etc.) which are not described at
each case and are not related  one to another.
\section{The advection-diffusion equation}We consider the advection in a random
velocity field ${\bf V}$ ( described in the Introduction)
 forced by a random source $f$\begin{equation}
 \partial_{\tau}\theta_{\tau}+{\bf V}\nabla
 \theta_{\tau}-\frac{\mu^{2}}{2}\triangle \theta_{\tau}=f
 \end{equation} where $\mu^{2}$ is the molecular diffusivity.
 If the random velocity ${\bf V}$ has correlation functions
 singular at small time then eq.(1) needs a careful
 interpretation. If the singularity of the velocity's covariance is of the form
 $\delta(t-t^{\prime})D({\bf x}-{\bf x}^{\prime})$ then there are two standard interpretations
 either Ito or Stratonovitch \cite{ikeda} \cite{simon}. The difference between
 them in eq.(1)is $\frac{1}{2}D({\bf 0})\nabla_{\bf z}^{2}\theta$. Hence,
 choosing one of them will  change only the diffusion constant.
 We choose the Stratonovitch interpretation throughout the paper
 and also in the Appendix B.

 First, let us consider ${\bf V}=0$ and $f=0$. Let
 $N$ be a (deterministic) solution of the heat equation  \begin{equation}
 \partial_{\tau}N_{\tau}-\frac{\mu^{2}}{2}\triangle N_{\tau}=0
 \end{equation}
 We expand  $\theta$ around the solution $N$ of the diffusion equation
 \begin{displaymath}
 \theta=T+N
 \end{displaymath}
 (if the mean value of ${\bf V}$ is zero then $T$ describes
 fluctuations of the temperature).
 From eq.(1)
\begin{equation}
 \partial_{\tau}T_{\tau}+{\bf V}\nabla
 T_{\tau}-\frac{\mu^{2}}{2}\triangle T_{\tau}=F
 \end{equation}
where
\begin{displaymath}
F=f-{\bf V}\nabla
 N_{\tau} \end{displaymath}

As the simplest example of a physical relevance we consider the
mean gradient \cite{gradient}\cite{arnovitz}
\begin{equation}
N=-{\bf g}{\bf X}
\end{equation}
where ${\bf g}$ is a constant vector. The mean gradient is a
stationary solution of the heat equation between two planes kept
at fixed temperatures. For such a static solution
\begin{equation}
F=f+{\bf V}{\bf g}
\end{equation}
We can see that even if $f=0$ then $F$ is non-trivial. This is a
frequent realization of an advection in experiments
\cite{gollub}\cite{biferale}. In such a case the source $F$ has
the same distribution as the velocity. A constant mean gradient is
distinguishing a direction in space. It breaks the rotational
symmetry. As a model we could consider ${\bf g}=(0,0,g_{z})$ and
${\bf V}=(0,0,V_{z})$.

We define the spectral measure $\rho$ of the temperature $T$ which
is directly measurable in experiments \cite{lesieur}
\begin{equation}\begin{array}{l}
\langle T_{\tau}({\bf x},{\bf z})T_{\tau}({\bf x}^{\prime},{\bf
z}^{\prime})\rangle-\langle T_{\tau}({\bf x},{\bf
z})\rangle\langle T_{\tau}({\bf x}^{\prime},{\bf
z}^{\prime})\rangle\cr=\int d{\bf k}d{\bf p}\exp(i{\bf k}({\bf
x}-{\bf x}^{\prime})+i{\bf p}({\bf z}-{\bf
z}^{\prime}))\rho_{\tau}({\bf k},{\bf p})
\end{array}\end{equation}We have

\begin{equation}
\begin{array}{l}
\int d{\bf x}(\langle \tilde{T}_{\tau}({\bf x},{\bf
p})\tilde{T}_{\tau}({\bf x}^{\prime},{\bf
p}^{\prime})\rangle-\langle \tilde{T}_{\tau}({\bf x},{\bf
p})\rangle\langle \tilde{T}_{\tau}({\bf x}^{\prime},{\bf
p}^{\prime})\rangle)\cr=\delta({\bf p}+{\bf
p}^{\prime})\rho_{\tau}({\bf 0},{\bf p}) \end{array}\end{equation}
and
\begin{displaymath}
\langle \tilde{T}_{\tau}({\bf x},{\bf p})\tilde{T}_{\tau}({\bf
x},{\bf p}^{\prime})\rangle-\langle \tilde{T}_{\tau}({\bf x},{\bf
p})\rangle\langle \tilde{T}_{\tau}({\bf x},{\bf
p})\rangle=\delta({\bf p}+{\bf p}^{\prime})\int d{\bf
k}\rho_{\tau}({\bf k},{\bf p})
\end{displaymath}
\begin{displaymath}
\langle T_{\tau}({\bf x},{\bf z})T_{\tau}({\bf x},{\bf
z})\rangle-\langle T_{\tau}({\bf x},{\bf z})\rangle\langle
T_{\tau}({\bf x},{\bf z})\rangle=\int d{\bf p}\int d{\bf
k}\rho_{\tau}({\bf k},{\bf p})
\end{displaymath}
When the spectral function has singularities at low momenta then
the Fourier transform in eq.(6) may need a careful definition in
the sense of generalized functions. Instead of the correlation
functions of $T_{\tau}({\bf x},{\bf y}) $ we could consider the
structure functions
\begin{displaymath}
{\cal G}^{(2n)}_{\tau}({\bf x},{\bf z})=\langle (T_{\tau}({\bf
0},{\bf 0})- \langle T_{\tau}({\bf 0},{\bf
0})\rangle-T_{\tau}({\bf x},{\bf z})+\langle T_{\tau}({\bf x},{\bf
z})\rangle)^{2n}\rangle
\end{displaymath}
For $ n=1$ we have
\begin{displaymath}
\begin{array}{l} {\cal G}^{(2)}_{\tau}({\bf x},{\bf z})=2\int d{\bf k}d
{\bf p}\rho_{\tau}({\bf k},{\bf p})(1-\exp(i{\bf k}{\bf x}+i{\bf
p}{\bf z})) \end{array}\end{displaymath} ${\cal G}^{(2)}_{\tau} $
scales in the same way as $\langle T T\rangle$ but has better
infrared behaviour.The structure functions ${\cal G}^{(2n)}$ are
expressed by the correlation functions of the Fourier transforms
of $T_{\tau}$.

 It can be seen
that the spectral measure $\rho$ of the temperature $T$ depends on
the spectral measure of the source $f$ and the scaling properties
of the random velocity field.

 \section{Gaussian model
of the shear  flow }

We decompose the fluid velocity
\begin{displaymath}
{\bf V}={\bf U}+{\bf v}
\end{displaymath}
into the mean value ${\bf U}$ and random fluctuations ${\bf v}$.
 We assume that the velocity ${\bf
v}$ is a  Gaussian Euclidean $R^{d}$ invariant random field with
the mean zero and the covariance
\begin{equation} \langle v_{j}(s,{\bf x})v_{k}(s^{\prime},{\bf
x}^{\prime})\rangle= G_{jk}(s-s^{\prime},{\bf x},{\bf x}^{\prime})
\end{equation}
where $j,k=d+1,...,D$. For the sake of simplicity of the arguments
we shall sometimes separate the time-dependence choosing $G$ of
the product form $\Gamma D$. If G is a decaying function of the
distance $\vert {\bf x}-{\bf x}^{\prime}\vert$ then a model of the
vector field  ${\bf v}$ can be determined by a translation
invariant G, e.g.,
\begin{equation}
G_{jk}(s-s^{\prime},{\bf x},{\bf
x}^{\prime})\equiv\delta_{jk}\Gamma(s-s^{\prime})D({\bf x}-{\bf
x}^{\prime})=\delta_{jk}\Gamma(s-s^{\prime})\int d{\bf p}
\exp(i{\bf p}( {\bf x}-{\bf x}^{\prime}))\tilde{D}({\bf p})
\end{equation}
where $\tilde{D}$ is a locally integrable function.

In a description of the turbulence we consider growing long range
correlations. In such a case $G$ cannot be translation invariant .
We consider a model with Euclidean $R^{d}$ invariant correlation
functions of ${\bf v}({\bf x})-{\bf v}({\bf x}^{\prime})$. Then
\begin{equation}
G_{jk}(s-s^{\prime},{\bf x},{\bf
x}^{\prime})=\delta_{jk}\Gamma(s-s^{\prime})(\vert {\bf
x}\vert^{2\beta}+\vert {\bf x}^{\prime}\vert^{2\beta}- \vert {\bf
x}-{\bf x}^{\prime}\vert^{2\beta})
\end{equation}
This $G$ is positive definite if $\Gamma$ is positive definite and
$0<\beta<1$ (the covariance (10) determines Levy's model
\cite{levy} of the Brownian motion depending on $d$-parameters ).
When $2\beta<2$ then the vector field ${\bf v}({\bf x})$ does not
satisfy the Lipschitz condition. In such a case we could expect
difficulties with the uniqueness of the flow and the uniqueness of
the solution of eq.(1) at $\mu=0$. Fortunately, a  definition of
the unique solution of eq.(1) in a weak probabilistic sense is
possible \cite{raymond}\cite{eiden}even without the Lipschitz
condition.

The source $f$ is an independent Gaussian field with the
covariance
\begin{equation}
\langle f(s,{\bf x},{\bf z})f(s^{\prime},{\bf x}^{\prime},{\bf
z}^{\prime})\rangle= M(s-s^{\prime},{\bf x}-{\bf x}^{\prime},{\bf
z}-{\bf z}^{\prime})
\end{equation}
We take the Fourier transform of eq.(3) in the ${\bf z}$ variable.
Then, this equation reads
\begin{equation}
\partial_{\tau}\tilde{T}_{\tau}({\bf x},{\bf p})+(i{\bf p}{\bf
V}(\tau,{\bf x})+\frac{\mu^{2}{\bf
p}^{2}}{2}-\frac{\mu^{2}}{2}\triangle_{\bf
x})\tilde{T}_{\tau}({\bf x},{\bf p})=\tilde{F}(\tau,{\bf x},{\bf
p})
\end{equation}
We apply the Feynman-Kac formula \cite{simon} in order to express
the solution of eq.(12) with the initial condition $T_{0}\in
L^{2}(dX)$ in the form (the uniqueness of the solution is
discussed in \cite{raymond}\cite{eiden})
\begin{equation}
\begin{array}{l}
\tilde{T}_{\tau}({\bf x},{\bf p})=\exp(-\frac{\mu^{2}{\bf
p}^{2}\tau}{2})E[\exp(-i{\bf p}\int_{0}^{\tau}{\bf V}(\tau -s,{\bf
x}+\mu{\bf b}(s))ds)\tilde{T}_{0}({\bf x}+\mu{\bf b}(\tau),{\bf
p})] +\cr \int_{0}^{\tau}dt\exp(-\frac{\mu^{2}{\bf
p}^{2}(\tau-t)}{2})E[\exp(-i{\bf p}\int_{0}^{\tau-t}{\bf V}(\tau
-s,{\bf x}+\mu{\bf b}(s))ds)\tilde{F}(t,{\bf x}+\mu{\bf
b}(\tau-t),{\bf p})]
\end{array}
\end{equation}
In eq.(13) $b_{j}$ ($j=1,2,...,d$) is the Brownian motion defined
as the Gaussian process with the covariance \cite{simon}
\begin{displaymath}
E[b_{j}(s)b_{k}(t)]=\delta_{jk}min(s,t) \end{displaymath}
 We are
interested in the equilibrium distribution of $T_{\tau}$, i.e.,in
the limit $\tau\rightarrow\infty$. When $\tau\rightarrow \infty$
and $T_{0}\in L^{2}(dX)$ then the first term in eq.(13) is
vanishing . For this reason we may set $T_{0}=0$ from the
beginning. The stationary solutions $N$ being harmonic functions
are not square integrable in $R^{D}$. Admitting such functions as
initial conditions we could regain the solution $N$ from eq.(13)
(with $F=0$ ). In particular, the mean gradient (4) comes from a
generalized function $\tilde{T}_{0}$ with its support concentrated
at ${\bf p}=0$.

Before discussing more general  correlations let us consider the
constant mean gradient (eqs.(4)-(5)) with $f=0$ and
 $F={\bf gV}$. Then, from eq.(13) (with $T_{0}=0$)\begin{equation}
\begin{array}{l}
\tilde{T}_{\tau}({\bf x},{\bf p})= \delta({\bf
p})E[\int_{0}^{\tau}dt{\bf gV} (t,{\bf x}+\mu {\bf b}(\tau-t))]
\end{array}
\end{equation}
We shall see that some properties of the general advection (3)
appear already at the level of the simple model (14). It follows
from eq.(14) that \begin{displaymath} \langle T_{\tau}({\bf
X})\rangle =\delta({\bf p})E[\int_{0}^{\tau}dt{\bf gU} (t,{\bf
x}+\mu {\bf b}(\tau-t))]
\end{displaymath}
and
\begin{displaymath}
\begin{array}{l}
\langle \tilde{T}_{\tau}({\bf x},{\bf p})\tilde{T}_{\tau}({\bf
x}^{\prime},{\bf p}^{\prime})\rangle-\langle \tilde{T}_{\tau}({\bf
x},{\bf p})\rangle\langle \tilde{T}_{\tau}({\bf x}^{\prime},{\bf
p}^{\prime})\rangle= \delta({\bf p})\delta({\bf
p}^{\prime})\cr\int_{0}^{\tau}dt\int_{0}^{\tau}dt^{\prime} E[{\bf
g}G(t-t^{\prime}, {\bf x}-{\bf x}^{\prime}+\mu {\bf
b}(\tau-t)-\mu{\bf b}^{\prime}(\tau-t^{\prime})){\bf g}]
\end{array}
\end{displaymath}
We calculate the integral over time. First, if the covariance $G$
is time-independent (a steady flow) then
\begin{equation}
\begin{array}{l}\langle T_{\tau}({\bf x},{\bf z})T_{\tau}({\bf
x}^{\prime},{\bf z}^{\prime})\rangle-\langle T_{\tau}({\bf
X})\rangle\langle T_{\tau}({\bf X}^{\prime})\rangle\cr
=4\mu^{-4}\int d{\bf k}\exp(i{\bf k}({\bf x}-{\bf
x}^{\prime})){\bf g}\tilde{G}({\bf k}){\bf g}\vert{\bf
k}\vert^{-4} (1-\exp(-\frac{\mu^{2}}{2}{\bf k}^{2}\tau))^{2}
\end{array}
\end{equation} Next, let us consider \begin{equation} G(t-t^{\prime},{\bf
x}-{\bf x}^{\prime})=\delta(t-t^{\prime})D({\bf x}-{\bf
x}^{\prime})
\end{equation}The covariance (16) does not have
 any physical foundations but the   virtue of the assumption (16)  is
the solubility of the model (3)
 \cite{kraichnan}(the Kraichnan model) in the sense that one can
 obtain a closed set of partial differential equations for the
 correlation functions (see the Appendix B).  In our simplified version (14)
\begin{equation}
\begin{array}{l}
\langle T_{\tau}({\bf x},{\bf z})T_{\tau}({\bf x}^{\prime},{\bf
z}^{\prime})\rangle-\langle T_{\tau}({\bf X})\rangle\langle
T_{\tau}({\bf X}^{\prime})\rangle\cr=\mu^{-2}\int d{\bf
k}\exp(i{\bf k}({\bf x}-{\bf x}^{\prime})){\bf g}\tilde{D}({\bf
k}){\bf g}\vert{\bf k}\vert^{-2} (1-\exp(-\mu^{2}{\bf
k}^{2}\tau))\end{array}
\end{equation}
If the ${\bf v}$ correlations are growing as in eq.(10) then the
expression (17) can be infrared divergent (especially at
$\tau=\infty$). In such a case we should rather consider
\begin{equation}\begin{array}{l} \langle (T_{\tau}({\bf 0},{\bf
0})-\langle T_{\tau}({\bf 0},{\bf 0})\rangle- T_{\tau}({\bf
x},{\bf z})+\langle T_{\tau}({\bf x},{\bf
z})\rangle)^{2}\rangle\cr=8\mu^{-4}\int d{\bf k}(1-\exp(i{\bf
k}{\bf x}))\tilde{G}({\bf k})\vert{\bf k}\vert^{-4}
(1-\exp(-\frac{\mu^{2}}{2}{\bf k}^{2}\tau))^{2}
\end{array}\end{equation} In general, let
\begin{equation}  G(t-t^{\prime},{\bf x}-{\bf x})
=\int d\omega d{\bf k}\tilde{G}(\omega,{\bf
k})\exp(i\omega(t-t^{\prime})+i{\bf k}({\bf x}-{\bf x}^{\prime}))
\end{equation}
then\begin{displaymath}\begin{array}{l} \langle
\tilde{T}_{\tau}({\bf x},{\bf p})\tilde{T}_{\tau}({\bf
x}^{\prime},{\bf p}^{\prime})\rangle-\langle \tilde{T}_{\tau}({\bf
x},{\bf p})\rangle\langle T_{\tau}({\bf x}^{\prime},{\bf
p}^{\prime})\rangle= \delta({\bf p})\delta({\bf
p}^{\prime})\cr\int_{0}^{\tau}dt\int_{0}^{\tau}dt^{\prime}\int
d{\bf k} {\bf g}\tilde{G}(t-t^{\prime},{\bf k}){\bf g}\exp(i{\bf
k}( {\bf x}-{\bf x}^{\prime})-\frac{1}{2}\mu^{2}{\bf k}^{2}
(2\tau-t-t^{\prime}))
\end{array}
\end{displaymath}After the time integration
\begin{equation}\begin{array}{l}
\langle T_{\tau}({\bf x},{\bf z})T_{\tau}({\bf x}^{\prime},{\bf
z}^{\prime})\rangle-\langle T_{\tau}(X)\rangle\langle
T_{\tau}(X^{\prime})\rangle=\int d{\bf k}d\omega\exp(i{\bf k}({\bf
x}-{\bf x}^{\prime})){\bf g}\tilde{G}(\omega,{\bf k}){\bf g}\cr
 (\frac{1}{4}\mu^{4}\vert{\bf k}\vert^{4}+\omega^{2})^{-1}\vert
 1-\exp(-\frac{1}{2}\mu^{2}{\bf k}^{2}\tau-i\omega\tau)\vert^{2}
 \end{array}
\end{equation}
We assume  that  $G$ is  scale invariant
\begin{equation} G(ct,\lambda {\bf x})=c^{-\alpha}
\lambda^{-2\gamma}G(t,{\bf x})
\end{equation}
($\alpha+\gamma<1$ if the time integral in eq.(14) is to be
finite). This  assumption has simple consequences for the heat
transport. It may be not exact in mathematical models. As an
example, for the shear flow solution of the Navier-Stokes equation
discussed in the Introduction if $C_{jl}(\omega,{\bf k})$ is the
spectral function of the force distribution ${\bf R}$ then the
spectral function of the stationary velocity distribution
(obtained as a solution of the Navier-Stokes equation with the
initial condition at $t_{0}$ and then letting $t_{0}\rightarrow
-\infty$)
 is
\begin{equation}\tilde{G}_{jl}(\omega,{\bf k})
=C_{jl}(\omega,{\bf k})\Big((\frac{\nu}{2}{\bf
k}^{2})^{2}+\omega^{2}\Big)^{-1}
\end{equation}
We must choose a specific $C$ in order to obtain a scale invariant
$\tilde{G}$.

 We can see from eqs.(15)-(20) that at finite $\tau$ the
large distance behavior of the temperature correlations is the
same as that of the velocity correlations because the behavior of
$\rho_{\tau}$ for small momenta does not change. However,if
$\langle {\bf v}({\bf x}){\bf v}({\bf 0})\rangle\simeq\vert {\bf
x}\vert^{2\beta}$ then  at $\tau=\infty$ for a steady flow we
obtain in eq.(18)\begin{displaymath} \langle (T_{\infty}({\bf
x},{\bf z})-\langle T_{\infty}({\bf x},{\bf
z})\rangle-T_{\infty}({\bf 0},{\bf 0})+\langle T_{\infty}({\bf
0},{\bf 0})\rangle)^{2}\rangle\simeq \vert {\bf x}\vert^{2\beta+4}
\end{displaymath}
 and for the Kraichnan model \cite{kraichnan}
\begin{displaymath}
\langle (T_{\infty}({\bf x},{\bf z})-\langle T_{\infty}({\bf
x},{\bf z})\rangle-T_{\infty}({\bf 0},{\bf 0})+\langle
T_{\infty}({\bf 0},{\bf 0})\rangle)^{2}\rangle\simeq \vert {\bf
x}\vert^{2\beta+2}
\end{displaymath}
in eq.(17) . For a general time dependent $G(t,{\bf x}) $ of the
form (19)  we shall have the $\vert{\bf x}
\vert^{2\beta-2\alpha+4}$ behavior of the structure functions
$S^{(2)}_{\infty}$ in eq.(20) if $G$ scales as in eq.(21) (
$\gamma=-\beta$). We can establish the behavior for large ${\bf
x}-{\bf x}^{\prime}$ by means of a change of variables in the
integrals (15)-(20) ${\bf k}=\tilde{{\bf k}}\vert{\bf x}-{\bf
x}^{\prime}\vert^{-1}$ and $\omega=\tilde{\omega}\vert{\bf x}-{\bf
x}^{\prime}\vert^{-2}$ and an estimate of the remainder. Note that
the long range correlations of the velocity field ($\gamma<0$)
lead to an increase of the temperature correlations.

\section{ Gaussian white noise source}
In this section we consider $F=f$ as a Gaussian random field
independent of ${\bf v}$. Estimates on the equilibrium
distribution are simplified if the sources at different times are
independent
\begin{equation}
\begin{array}{l}
M(t-t^{\prime},{\bf x}-{\bf x}^{\prime},{\bf z}-{\bf z}^{\prime})
=\delta(t-t^{\prime})m({\bf x}-{\bf x}^{\prime},{\bf z}-{\bf
z}^{\prime})
\end{array}
\end{equation}
We assume the form (23) of $M$ as a technical simplification. This
is a mathematical idealization still justified by an application
of physical sources of heat (as heat injections independent at
each time).

For a lower bound we need an assumption that the dependence on
${\bf x}-{\bf x}^{\prime}$ is of the form of the Laplace transform
(such an assumption includes  the scale invariant distributions
$m$ which do not increase at large distances) either in the form
\begin{equation}
\begin{array}{l} m({\bf x}-{\bf x}^{\prime},{\bf z}-{\bf
z}^{\prime})\equiv m_{1}({\bf x}-{\bf x}^{\prime})m_{0}({\bf
z}-{\bf z}^{\prime})\cr =\int d{\bf k}d{\bf p}\exp(i{\bf k}({\bf
x}-{\bf x}^{\prime})+i{\bf p}({\bf z}-{\bf
z}^{\prime}))\tilde{m}_{1}({\bf k})\tilde{m}_{0}({\bf p})\cr
 =
\int_{0}^{\infty}da \nu_{1}(a)\exp(-a\vert {\bf x}-{\bf
x}^{\prime}\vert^{2})m_{0}({\bf z}-{\bf z}^{\prime})
\end{array}
\end{equation}
or in the Euclidean invariant way
\begin{equation}
\begin{array}{l} m({\bf x}-{\bf x}^{\prime},{\bf z}-{\bf
z}^{\prime}) = \int_{0}^{\infty}da \nu(a)\exp(-a(\vert {\bf
x}-{\bf x}^{\prime}\vert^{2}+\vert {\bf z}-{\bf
z}^{\prime}\vert^{2})) \cr \equiv \int_{0}^{\infty}da\int d{\bf
p}\exp(i{\bf p}({\bf z}-{\bf z}^{\prime}))\exp(-a\vert {\bf
x}-{\bf x}^{\prime}\vert^{2})\nu(a,{\bf p})
\end{array}
\end{equation}
In eqs.(24)-(25)  $\nu_{1}$ and $\nu$ are non-negative functions.

 ${\bf v}$  in eq.(13) enters $T_{\tau}$ in the form
\begin{displaymath}
\exp(i{\bf v}({\bf J}))
\end{displaymath}
where
\begin{displaymath}
{\bf v}({\bf J})=\int d{\bf u}\int_{0}^{\tau} ds {\bf v}(s,{\bf
u}){\bf J}(s,{\bf u})
\end{displaymath}
with
\begin{displaymath} {\bf J}(s,{\bf u})=-\theta(s){\bf
p}\delta({\bf u}-{\bf x}-\mu{\bf b}(\tau-s))\end{displaymath}  It
follows that the expectation values of  $n$ products of $T_{\tau}$
are expressed by
\begin{displaymath}
\langle \exp(i{\bf v}({\bf J}_{n}))\rangle=S({\bf J}_{n})
\end{displaymath}
where $S({\bf J})$ is the characteristic function of the random
field ${\bf v}$. For a Gaussian random field
\begin{equation}
S({\bf J})=\exp(-\frac{1}{2}{\bf J}G{\bf J})
\end{equation}
Let us note that because of the translation invariance in the
${\bf z}$ variable of the source $f$  we have a conservation of
momenta
\begin{equation}
\langle \tilde{T}_{\tau}({\bf x}_{1},{\bf
p}_{1}).....\tilde{T}_{\tau}({\bf x}_{n},{\bf p}_{n})\rangle
=\delta({\bf p}_{1}+...+{\bf p}_{n}){\cal H}\end{equation} The
correlation functions (27) are expressed by the characteristic
function (26) with ${\bf J}_{n}$ satisfying the condition (for
$n>1$)
\begin{equation} \int{\bf J}_{n}(s,{\bf u})d{\bf u}=0
\end{equation}
It follows that in the Gaussian case with the covariance (10) the
part   of $G$ which is not translation invariant does not
contribute to the correlation functions.

We calculate the equal time expectation values of $T_{\tau}$ (
eq.(13)with the zero initial condition) under the assumption that
the random fields $f$ and ${\bf v}$ are independent
\begin{equation}
\begin{array}{l}
\langle \tilde{T}_{\tau}({\bf x},{\bf p})\tilde{T}_{\tau}({\bf
x}^{\prime},{\bf p}^{\prime})\rangle=  \delta({\bf p}+{\bf
p}^{\prime})\int_{0}^{\tau}dt \exp(-\mu^{2}{\bf p}^{2}(\tau-t))
\cr E[\exp(-i{\bf p}\int_{0}^{\tau-t}{\bf U}(\tau -s,{\bf
x}+\mu{\bf b}(s))ds)\cr \tilde{m}( {\bf x}-{\bf x}^{\prime}+\mu
{\bf b}(\tau-t)-\mu{\bf b}^{\prime}(\tau-t),{\bf p})S({\bf
J}_{2})]
\end{array}
\end{equation}
where \begin{displaymath} {\bf J}_{2}({\bf u})= {\bf p}\theta
(s)\delta({\bf u}-{\bf x}-{\bf b}(\tau-s))-{\bf p}\theta
(s)\delta({\bf u}-{\bf x}^{\prime}-{\bf b}^{\prime}(\tau-s))
\end{displaymath} For the Gaussian field (26)

\begin{equation}
\begin{array}{l}

S({\bf
J}_{2})=\exp\Big(-\frac{1}{2}\int_{0}^{\tau-t}\int_{0}^{\tau-t}dsds^{\prime}
{\bf p} G_{0}(s-s^{\prime},\mu{\bf b}(s)-\mu{\bf
b}(s^{\prime})){\bf p} \cr -
\frac{1}{2}\int_{0}^{\tau-t}\int_{0}^{\tau-t}dsds^{\prime} {\bf p}
G_{0}(s-s^{\prime},\mu{\bf b}^{\prime}(s)-\mu{\bf
b}^{\prime}(s^{\prime})){\bf p} \cr +\int_{0}^{\tau
-t}\int_{0}^{\tau-t}dsds^{\prime} {\bf p} G_{0}(s-s^{\prime},{\bf
x}-{\bf x}^{\prime}+\mu{\bf b}(s)-\mu{\bf
b}^{\prime}(s^{\prime})){\bf p} \Big)
\end{array}
\end{equation}
where $G_{0}$ is the translation invariant part of $G$.

If \begin{equation} \vert \tilde{m}( {\bf x},{\bf p})\vert\leq
K\vert\tilde{m}_{0}\vert({\bf p})
\end{equation}
then from  $\vert S({\bf J})\vert\leq 1$ there follows the
bound\begin{equation} \vert\langle \tilde{T}_{\tau}({\bf x},{\bf
p})\tilde{T}_{\tau}({\bf x}^{\prime},{\bf p}^{\prime})\rangle\vert
\leq K\delta({\bf p}+{\bf p}^{\prime})\vert\tilde{m}_{0}\vert({\bf
p})\mu^{-2}{\bf p}^{-2}
 (1-\exp(-\mu^{2}{\bf p}^{2}\tau))\end{equation}
For a small ${\bf p}$ and a finite $\tau$ the correlations (32)
are bounded by $\tau\vert\tilde{m}_{0}\vert({\bf p})$ whereas at
$\tau=\infty$ by $\vert\tilde{m}_{0}\vert({\bf p}){\bf p}^{-2}$.

Next, we apply the scale invariance of the Brownian motion
\begin{equation}
{\bf b}(at)=\sqrt{a}{\bf b}(t)
\end{equation}
in eq.(29).We write $s=(\tau-t)\sigma$.
 Then, using the scaling properties (21) and (33) and denoting by $G_{0}$
 the translation invariant part of $G$ we can rewrite eqs.(29)-(30) in
the form
\begin{equation}
\begin{array}{l}
\langle \tilde{T}_{\tau}({\bf x},{\bf p})\tilde{T}_{\tau}({\bf
x}^{\prime},{\bf p}^{\prime})\rangle=  \delta({\bf p}+{\bf
p}^{\prime})\int_{0}^{\tau}dt \exp(-\mu^{2}{\bf p}^{2}(\tau-t))\cr
E[\exp(-i{\bf p}\int_{0}^{\tau-t}{\bf U}(\tau -s,{\bf x}+\mu{\bf
b}(s))ds)\tilde{m}({\bf x}-{\bf x}^{\prime}+\mu\sqrt{\tau-t}{\bf
b}(1)-\mu\sqrt{\tau-t}{\bf b}^{\prime}(1),{\bf p})\cr
\exp\Big(-\frac{1}{2}(\tau-t)^{2-\alpha-\gamma}
\int_{0}^{1}\int_{0}^{1}d\sigma d\sigma^{\prime} {\bf p}
G_{0}(\sigma-\sigma^{\prime},\mu{\bf b}(\sigma)-\mu{\bf
b}(\sigma^{\prime})){\bf p} \cr
-\frac{1}{2}(\tau-t)^{2-\alpha-\gamma}\int_{0}^{1}\int_{0}^{1}d\sigma
d\sigma^{\prime} {\bf p} G_{0}(\sigma-\sigma^{\prime},\mu{\bf
b}^{\prime}(\sigma)-\mu{\bf b}^{\prime}(\sigma^{\prime})){\bf p}
\cr +(\tau-t)^{2-\alpha-\gamma} \int_{0}^{1}\int_{0}^{1}d\sigma
d\sigma^{\prime} {\bf p}
G_{0}(\sigma-\sigma^{\prime},(\tau-t)^{-\frac{1}{2}}({\bf x}-{\bf
x}^{\prime})+\mu{\bf b}(\sigma)-\mu{\bf
b}^{\prime}(\sigma^{\prime})){\bf p}\Big)]
\end{array}
\end{equation}
For the Kraichnan model \cite{kraichnan}
$\Gamma(s-s^{\prime})=\delta(s-s^{\prime})$ in eqs.(9)-(10),then
$\alpha=1$ in eq.(22) and the formula (34) reads

\begin{equation}
\begin{array}{l}
\langle \tilde{T}_{\tau}({\bf x},{\bf p})\tilde{T}_{\tau}({\bf
x}^{\prime},{\bf p}^{\prime})\rangle= \delta({\bf p}+{\bf
p}^{\prime})\int_{0}^{\tau}dt \exp(-\mu^{2}{\bf p}^{2}(\tau-t))\cr
E[\exp(-i{\bf p}\int_{0}^{\tau-t}{\bf U}(\tau -s,{\bf x}+\mu{\bf
b}(s))ds)\tilde{m}({\bf x}-{\bf x}^{\prime}+\mu\sqrt{\tau-t}{\bf
b}(1)-\mu\sqrt{\tau-t}{\bf b}^{\prime}(1),{\bf p})\cr
\exp\Big(-(\tau-t)^{1-\gamma} {\bf p} D({\bf 0}){\bf p}
+(\tau-t)^{1-\gamma}\int_{0}^{1}d\sigma {\bf p}
D((\tau-t)^{-\frac{1}{2}}({\bf x}-{\bf x}^{\prime})+\mu{\bf
b}(\sigma)-\mu{\bf b}^{\prime}(\sigma)){\bf p}\Big)]
\end{array}
\end{equation}

\section{Jensen inequalities for the temperature correlations}
  We are going to estimate
the spectral measure (6)-(7) by an application of the Jensen
inequality. We can obtain an upper bound on the correlation
functions applying the Jensen inequality ($\int d\nu\exp f\geq
\exp(\int d\nu f$) if $\int d\nu=1$) \cite{jensen} to the time
integral in eqs.(29)-(30)
\begin{equation}
\begin{array}{l}
\vert\langle \tilde{T}_{\tau}({\bf x},{\bf
p})\tilde{T}_{\tau}({\bf x}^{\prime},{\bf p}^{\prime})\rangle
\vert\leq \cr 2\delta({\bf p}+{\bf
p}^{\prime})\int_{0}^{\tau}dr\int_{0}^{1}d\sigma
\int_{0}^{\sigma}d\sigma^{\prime}\exp(-\mu^{2}{\bf p}^{2}r) \cr
E[\vert\tilde{m}( {\bf x}-{\bf x}^{\prime}+\mu\sqrt{r}{\bf
b}(1)-\mu\sqrt{r}{\bf b}^{\prime}(1),{\bf p})\vert
\cr\exp\Big(-\frac{1}{2}r^{2-\alpha-\gamma} {\bf p}
G_{0}(\sigma-\sigma^{\prime},\mu{\bf b}(\sigma)-\mu{\bf
b}(\sigma^{\prime})){\bf p} \cr -\frac{1}{2}r^{2-\alpha-\gamma}
{\bf p} G_{0}(\sigma-\sigma^{\prime},\mu{\bf
b}^{\prime}(\sigma)-\mu{\bf b}^{\prime}(\sigma^{\prime})){\bf p}
\cr +r^{2-\alpha-\gamma} {\bf p}
G_{0}(\sigma-\sigma^{\prime},r^{-\frac{1}{2}}({\bf x}-{\bf
x}^{\prime})+\mu{\bf b}(\sigma)-\mu{\bf
b}^{\prime}(\sigma^{\prime})){\bf p}\Big)]
\end{array}
\end{equation}
Let $p(s,{\bf u};t,{\bf w})$ be the transition function for the
Brownian motion to pass from ${\bf u}$ at time $s$ to ${\bf w}$ at
time $t$. Then, the expectation value (36) reads

\begin{equation}
\begin{array}{l}
\vert\langle \tilde{T}_{\tau}({\bf x},{\bf
p})\tilde{T}_{\tau}({\bf x}^{\prime},{\bf
p}^{\prime})\rangle\vert\leq \cr 2\delta({\bf p}+{\bf
p}^{\prime})\int d{\bf u}d{\bf u}^{\prime}d{\bf w}d{\bf
w}^{\prime}\int_{0}^{\tau}dr\int_{0}^{1}d\sigma
\int_{0}^{\sigma}d\sigma^{\prime}\exp(-\mu^{2}{\bf p}^{2}r) \cr
p(0,{\bf 0};\sigma^{\prime},{\bf u}) p(\sigma^{\prime},{\bf
u};\sigma,{\bf w}) p(\sigma,{\bf w};1,{\bf z}) p(0,{\bf
0};\sigma^{\prime},{\bf u}^{\prime}) p(\sigma^{\prime},{\bf
u}^{\prime};\sigma,{\bf w}^{\prime}) p(\sigma,{\bf
w}^{\prime};1,{\bf z}^{\prime}) \cr \vert\tilde{m}( {\bf x}-{\bf
x}^{\prime}+\mu\sqrt{r}{\bf z}-\mu\sqrt{r}{\bf z}^{\prime},{\bf
p})\vert \cr\exp\Big(-\frac{1}{2}r^{2-\alpha-\gamma} {\bf p}
G_{0}(\sigma-\sigma^{\prime},\mu{\bf w}-\mu{\bf u}){\bf p}
-\frac{1}{2}r^{2-\alpha-\gamma} {\bf p}
G_{0}(\sigma-\sigma^{\prime},\mu{\bf w}^{\prime}-\mu{\bf
u}^{\prime}){\bf p} \cr +\frac{1}{2}r^{2-\alpha-\gamma} {\bf p}
G_{0}(\sigma-\sigma^{\prime},r^{-\frac{1}{2}}({\bf x}-{\bf
x}^{\prime})+\mu{\bf w}-\mu{\bf u}^{\prime}){\bf p} \cr
+\frac{1}{2}r^{2-\alpha-\gamma} {\bf p}
G_{0}(\sigma-\sigma^{\prime},r^{-\frac{1}{2}}({\bf x}-{\bf
x}^{\prime})+\mu{\bf w}^{\prime}-\mu{\bf u}){\bf p}\Big)
\end{array}
\end{equation}
Till now we have kept the mean velocity ${\bf U}$ as an arbitrary
non zero function.We can obtain a lower bound only if
\begin{displaymath} {\bf U}=0
\end{displaymath}
As claimed by some authors (see ,e.g.,the standard text-book
\cite{landau}) the mean velocity does not play any essential role
in turbulence. So, setting it equal to zero we do not lose much.
Moreover, for the lower bound we must assume $m$ of the form (24)
(or (25)) with $\tilde{m}_{0}({\bf p})\geq 0$. Then, we can apply
the Jensen inequality to the expectation value over the Brownian
motion
\begin{equation}
\begin{array}{l}
\langle \tilde{T}_{\tau}({\bf x},{\bf p})\tilde{T}_{\tau}({\bf
x}^{\prime},{\bf p}^{\prime})\rangle\geq \cr \delta({\bf p}+{\bf
p}^{\prime})\int_{0}^{\tau}dr \exp(-\mu^{2}{\bf
p}^{2}r)\int_{0}^{\infty} da \nu_{1}(a)\cr\tilde{m}_{0}({\bf p})
\exp E\Big[-\frac{1}{2}r^{2-\alpha-\gamma}
\int_{0}^{1}\int_{0}^{1}d\sigma d\sigma^{\prime} {\bf p}
G_{0}(\sigma-\sigma^{\prime},\mu{\bf b}(\sigma)-\mu{\bf
b}(\sigma^{\prime})){\bf p} \cr
-\frac{1}{2}r^{2-\alpha-\gamma}\int_{0}^{1}\int_{0}^{1}d\sigma
d\sigma^{\prime} {\bf p} G_{0}(\sigma-\sigma^{\prime},\mu{\bf
b}^{\prime}(\sigma)-\mu{\bf b}^{\prime}(\sigma^{\prime})){\bf p}
\cr +r^{2-\alpha-\gamma} \int_{0}^{1}\int_{0}^{1}d\sigma
d\sigma^{\prime} {\bf p}
G_{0}(\sigma-\sigma^{\prime},r^{-\frac{1}{2}}({\bf x}-{\bf
x}^{\prime})+\mu{\bf b}(\sigma)-\mu{\bf
b}^{\prime}(\sigma^{\prime})){\bf p}\cr-a\vert {\bf x}-{\bf
x}^{\prime}+\mu\sqrt{r}{\bf b}(1)-\mu\sqrt{r}{\bf
b}^{\prime}(1)\vert^{2}\Big]
\end{array}
\end{equation}
For $m$ of the form (25) the inequality (38) reads

\begin{equation}
\begin{array}{l}
\langle \tilde{T}_{\tau}({\bf x},{\bf p})\tilde{T}_{\tau}({\bf
x}^{\prime},{\bf p}^{\prime})\rangle\geq \cr \delta({\bf p}+{\bf
p}^{\prime})\int_{0}^{\tau}dr \exp(-\mu^{2}{\bf
p}^{2}r)\int_{0}^{\infty} da \nu(a,{\bf p})\cr \exp
E\Big[-\frac{1}{2}r^{2-\alpha-\gamma}
\int_{0}^{1}\int_{0}^{1}d\sigma d\sigma^{\prime} {\bf p}
G_{0}(\sigma-\sigma^{\prime},\mu{\bf b}(\sigma)-\mu{\bf
b}(\sigma^{\prime})){\bf p} \cr
-\frac{1}{2}r^{2-\alpha-\gamma}\int_{0}^{1}\int_{0}^{1}d\sigma
d\sigma^{\prime} {\bf p} G_{0}(\sigma-\sigma^{\prime},\mu{\bf
b}^{\prime}(\sigma)-\mu{\bf b}^{\prime}(\sigma^{\prime})){\bf p}
\cr +r^{2-\alpha-\gamma} \int_{0}^{1}\int_{0}^{1}d\sigma
d\sigma^{\prime} {\bf p}
G_{0}(\sigma-\sigma^{\prime},r^{-\frac{1}{2}}({\bf x}-{\bf
x}^{\prime})+\mu{\bf b}(\sigma)-\mu{\bf
b}^{\prime}(\sigma^{\prime})){\bf p}\cr-a\vert {\bf x}-{\bf
x}^{\prime}+\mu\sqrt{r}{\bf b}(1)-\mu\sqrt{r}{\bf
b}^{\prime}(1)\vert^{2}\Big]
\end{array}
\end{equation}
The correlation functions (36)-(39) in general will essentially
depend on the source distribution $m$. We consider  $m$ such
that:i) $m_{1}$ is bounded  from above by a constant (eq.(31)) and
in addition ii) $m_{1}({\bf x})$ is decreasing like a power
$2\Omega$ of $\vert {\bf x}\vert$.
 From eq.(32) it
follows that under the assumption (31) the limit $\tau\rightarrow
\infty$ exists. We wish to estimate the correlation functions at
$\tau=\infty$ under various conditions on $m_{1}({\bf x})$ . Using
the inequality (for $A\geq 0$)
 \begin{displaymath}
 2\exp(-\mu^{2}{\bf p}^{2}r-A({\bf x}-{\bf x}^{\prime},{\bf b})r^{2-\alpha-\gamma}{\bf p}^{2})
 \leq \exp(-\mu^{2}{\bf p}^{2}r)+
  \exp(-A({\bf x}-{\bf x}^{\prime},{\bf b}) r^{2-\alpha-\gamma}{\bf p}^{2})
 \end{displaymath}
 and a change of variables in the $r$-integral in eqs.(36)-(37)
  $r=t\vert{\bf p}\vert^{-\frac{2}{2-\alpha-\gamma}}$ we
 obtain ( when $m_{1}$ is a bounded function (31))
\begin{equation}
\begin{array}{l}\langle \tilde{T}_{\infty}({\bf x},{\bf
p})\tilde{T}_{\infty}({\bf x},{\bf p}^{\prime})\rangle\cr\leq
\delta({\bf p}+{\bf p}^{\prime}) \vert\tilde{m}_{0}\vert({\bf
p})\Big(c_{1}\theta(\vert {\bf p}\vert -\frac{1}{\mu})\vert {\bf
p}\vert^{-2}+c_{2}\theta(\frac{1}{\mu}-\vert {\bf p}\vert )\vert
{\bf p}\vert^{-\frac{2}{2-\alpha-\gamma}}\Big)
\end{array}
\end{equation}
where on the rhs of eq.(36) after an integral over $r$ (which can
be performed by a change of variables) we  obtain a function
$\vert A({\bf x}-{\bf x}^{\prime},{\bf
b})\vert^{-\frac{1}{2-\alpha-\gamma}}$ (where ${\bf b}$ depends on
$\sigma$ and $\sigma^{\prime}$) whose expectation value is
expressed by the rhs of eq.(37). This is an integrable function of
${\bf u},{\bf w},{\bf z},{\bf u}^{\prime},{\bf w}^{\prime}$ and
${\bf z}^{\prime}$. Hence, it can be bounded by a constant
$c_{2}$. Under a stronger assumption that
\begin{equation}
\int d{\bf x}m_{1}({\bf x})<\infty
\end{equation}
from eq.(37) we obtain in a similar way the bound
\begin{equation}
\begin{array}{l} \int d{\bf
x}\langle \tilde{T}_{\infty}({\bf x},{\bf
p})\tilde{T}_{\infty}({\bf x}^{\prime},{\bf
p}^{\prime})\rangle\cr\leq \delta({\bf p}+{\bf p}^{\prime})
\vert\tilde{m}_{0}\vert({\bf p})\Big(c_{3}\theta(\vert {\bf
p}\vert -\frac{1}{\mu})\vert {\bf
p}\vert^{-2}+c_{4}\theta(\frac{1}{\mu}-\vert {\bf p}\vert )\vert
{\bf p}\vert^{-\frac{2}{2-\alpha-\gamma}}\Big)
\end{array}
\end{equation}
This is a bound on the spectral measure on the rhs of eq.(7).

We wish to estimate the dependence of the correlation functions
(34) on ${\bf x}-{\bf x}^{\prime}$ in a more explicit form. Note
that if the velocity correlations are defined by eq.(9) where
$\tilde{D}({\bf k})$ is an integrable function then on the basis
of the Lebesgue lemma $G$ is vanishing at large $\vert {\bf
x}-{\bf x}^{\prime}\vert$. In such a case the term depending on
${\bf x}-{\bf x}^{\prime}$ in the exponential on the rhs of
eq.(34) can be neglected. If $m$ is in addition a slowly varying
function of ${\bf x}-{\bf x}^{\prime}$ then
\begin{equation} \langle \tilde{T}_{\tau}({\bf x},{\bf
p})\tilde{T}_{\tau}({\bf x}^{\prime},{\bf
p}^{\prime})\rangle\simeq\langle \tilde{T}_{\tau}({\bf x},{\bf
p})\tilde{T}_{\tau}({\bf x},{\bf p}^{\prime})\rangle
\end{equation}
There remains to discuss the turbulent flow (10). We are unable to
prove precise upper bounds for  large $\vert {\bf x}-{\bf
x}^{\prime}\vert$ and general $\beta$. However, if $0<2\beta<1$
and $d=1$ then $g({\bf x})=-\vert {\bf x}\vert^{2\beta}$ is a
convex function \cite{jensen}
\begin{displaymath}
g(\frac{1}{2}({\bf x}+{\bf y}))\leq \frac{1}{2}g({\bf x})
+\frac{1}{2}g({\bf y})
\end{displaymath}
As a consequence
\begin{displaymath}
\begin{array}{l}
\exp\Big(-r^{2-\alpha+\beta} {\bf p}^{2}
\Gamma(\sigma-\sigma^{\prime})\vert r^{-\frac{1}{2}}({\bf x}-{\bf
x}^{\prime})+\mu{\bf b}(\sigma)-\mu{\bf
b}^{\prime}(\sigma^{\prime})\vert^{2\beta}\Big)\cr \leq
\exp\Big(-\frac{1}{2}r^{2-\alpha+\beta} {\bf p}^{2}
\Gamma(\sigma-\sigma^{\prime})\vert 2 r^{-\frac{1}{2}}({\bf
x}-{\bf x}^{\prime})\vert^{2\beta}-\frac{1}{2}\vert 2\mu{\bf
b}(\sigma)-2\mu{\bf
b}^{\prime}(\sigma^{\prime})\vert^{2\beta}\Big)
\end{array}
\end{displaymath}
Hence, under the assumption (31) (after the $r$-integration) the
inequalities (36)-(37) at $\tau=\infty$ for $0<2\beta<1$ read
\begin{equation}
\begin{array}{l}
\langle \tilde{T}_{\infty}({\bf x},{\bf p})\tilde{T}_{\infty}({\bf
x}^{\prime},{\bf p}^{\prime})\rangle\cr
 \leq K \vert {\bf p}\vert^{-\frac{2}{2-\alpha}}\vert\tilde{m}_{0}\vert({\bf p})
 \vert {\bf x}-{\bf x}^{\prime}\vert^{-\frac{2\beta}{2-\alpha}}
 \end{array}
 \end{equation}
 We expect the inequality (44) to hold true in general
 (under the assumption (31))for large
 $\vert {\bf x}-{\bf x}^{\prime}\vert$ because we obtain
 such a behavior of the two-point function if in a formal way
 we take the limit $\vert {\bf x}-{\bf x}^{\prime}\vert\rightarrow \infty$
 in eq.(34) neglecting terms of order
 $\vert {\bf x}-{\bf x}^{\prime}\vert^{-1}$.

We discuss now the Jensen inequality (38) for the lower bound.
 It is sufficient to calculate the expectation value in
the exponential (38).
 First, in
the Kraichnan model (35)  for the term $-W$ in the exponential
appearing in eq.(38) we obtain
\begin{equation}
\begin{array}{l}
\exp(-W({\bf x}-{\bf x}^{\prime}))=\exp\Big(-r^{1-\gamma}\int
d{\bf k}{\bf p}\tilde{D}\left({\bf k}\right){\bf p}
\cr\left(1-\mu^{-2}{\bf k}^{-2}\exp\left(i{\bf
k}r^{-\frac{1}{2}}\left({\bf x}- {\bf x}^{\prime}\right)\right)
\left(1-\exp\left(-\mu^{2}{\bf k }^{2}\right)\right)\right)\Big)
\end{array}
\end{equation}
It is easy to see that

\begin{equation}
\begin{array}{l}
\exp(-W({\bf 0}))=\exp\Big(-r^{1-\gamma}\int d{\bf k}{\bf
p}\tilde{D}\left({\bf k}\right){\bf p} \cr\left(1-\mu^{-2}{\bf
k}^{-2} \left(1-\exp\left(-\mu^{2}{\bf k
}^{2}\right)\right)\right)\Big)\geq \exp(-cr^{1-\gamma}{\bf
p}^{2})
\end{array}
\end{equation}
under the assumptions that $ {\bf p}\tilde{D}{\bf p}\geq
\vert\tilde{D}\vert {\bf p}^{2}$, $\int d{\bf
k}\vert\tilde{D}\vert({\bf k})\theta (\vert {\bf k}\vert
-\frac{1}{\mu})<\infty$ and
\begin{displaymath}
\int d{\bf k}\vert\tilde{D}\vert({\bf k}){\bf k}^{2}\theta
(\frac{1}{\mu}-\vert {\bf k}\vert)<\infty.
\end{displaymath}
In such a case we can take the limit $\tau\rightarrow \infty$. In
this limit
\begin{equation}
\begin{array}{l}\langle \tilde{T}_{\infty}({\bf x},{\bf
p})\tilde{T}_{\infty}({\bf x},{\bf p}^{\prime})\rangle\geq
\delta({\bf p}+{\bf p}^{\prime})\tilde{m}_{0}({\bf
p})\int_{0}^{\infty}dr \int da \nu_{1}(a)\cr \exp(-\mu^{2}{\bf
p}^{2}r-c {\bf p}^{2}r^{1-\gamma}-2a\mu^{2} r)
\cr=\tilde{m}_{0}({\bf p})\delta({\bf p}+{\bf
p}^{\prime})\int_{0}^{\infty}drm_{1}( \mu\sqrt{2
r})\exp(-\mu^{2}{\bf p}^{2}r-c {\bf p}^{2}r^{1-\gamma})
\end{array}
\end{equation}
The behavior of the integral (47) depends on the behavior of the
source correlations $m_{1}$ as a function of $\vert {\bf x}-{\bf
x}^{\prime}\vert$. If
\begin{equation} m_{1}(\mu\sqrt{2 r})\geq K
\end{equation}
then
\begin{equation}\langle \tilde{T}_{\infty}({\bf x},{\bf
p})\tilde{T}_{\infty}({\bf x},{\bf p}^{\prime})\rangle\geq
\delta({\bf p}+{\bf p}^{\prime})\tilde{m}_{0}({\bf p})
\Big(c_{5}\theta(\vert {\bf p}\vert -\frac{1}{\mu})\vert{\bf
p}\vert^{-2}+c_{6}\theta(\frac{1}{\mu}-\vert {\bf p}\vert )\vert
{\bf p}\vert^{-\frac{2}{1-\gamma}}\Big)
\end{equation}
This lower bound coincides with the upper bound (40) (where
$\alpha=1$). If $m_{1}$ satisfies  a stronger condition (
$\Omega<1$)\begin{equation} m_{1}(\vert {\bf x}\vert)\geq K \vert
{\bf x}\vert^{ -2\Omega}
\end{equation}
( $\Omega\geq 0$ if it is to be of  the form (24),i.e.,
$\nu_{1}(a)\geq Ka^{\Omega-1}$) then
\begin{equation}\langle \tilde{T}_{\infty}({\bf x},{\bf
p})\tilde{T}_{\infty}({\bf x},{\bf p}^{\prime})\rangle\geq
\delta({\bf p}+{\bf p}^{\prime})\tilde{m}_{0}({\bf p})
\Big(c_{7}\theta(\vert {\bf p}\vert -\frac{1}{\mu})\vert{\bf
p}\vert^{-2+2\Omega}+c_{8}\theta(\frac{1}{\mu}-\vert {\bf p}\vert
)\vert {\bf p}\vert^{-\frac{2-2\Omega}{1-\gamma}}\Big)
\end{equation}
 The inequality (51) results from the following estimate (for
 $\alpha+\gamma<1$)
\begin{equation}
\begin{array}{l}
\int_{0}^{\infty}dr r^{-\Omega}\exp(-\mu^{2}{\bf p}^{2}r-c {\bf
p}^{2}r^{2-\alpha-\gamma})\cr=\int_{0}^{1}drr^{-\Omega}
\exp(-\mu^{2}{\bf p}^{2}r-c{\bf p}^{2}r^{2-\alpha-\gamma})
 +\int_{1}^{\infty}drr^{-\Omega}\exp(-\mu^{2}{\bf p}^{2}r-c{\bf
p}^{2}r^{2-\alpha-\gamma}) \cr\geq \int_{0}^{1}drr^{-\Omega}
\exp(-(\mu^{2}{\bf p}^{2}+c{\bf p}^{2})r)
+\int_{1}^{\infty}drr^{-\Omega}\exp(-(\mu^{2}{\bf p}^{2}+c{\bf
p}^{2})r^{2-\alpha-\gamma}) \cr = \vert{\bf
p}\vert^{-2+2\Omega}\int_{0}^{{\bf
p}^{2}}t^{-\Omega}\exp(-(\mu^{2}+c)t)dt + \vert{\bf
p}\vert^{-\frac{2-2\Omega}{2-\alpha-\gamma}}\int_{a({\bf
p})}^{\infty}t^{-\Omega}\exp(-(\mu^{2}+c)t^{2-\alpha-\gamma})dt
\end{array}
\end{equation}
where $a({\bf p})=\vert{\bf p}\vert^{\frac{2}{2-\alpha-\gamma}}$
and  $\alpha=1$ in application to eq.(47).

 Next, we wish to estimate the behavior of the
temperature correlations at large ${\bf x}-{\bf x}^{\prime}$ in
the turbulent case (10) when $\gamma=-\beta<0$ (if $\gamma>0$ and
$m_{1}$ is a bounded function then the temperature correlations
are bounded from below and from above as functions of ${\bf
x}-{\bf x}^{\prime}$ ,eq.(43)). First, we consider the  Kraichnan
model (35) ($\Gamma(s-s^{\prime})=\delta(s-s^{\prime})$ in
eq.(10)) with the mean velocity ${\bf U}=0$ and
\begin{equation}
\tilde{D}({\bf k})\simeq \vert{\bf k}\vert^{-d+2\gamma}
\end{equation}
The integral in eq.(45) is convergent for large ${\bf k}$ if
$\gamma<0$ and for small ${\bf k}$ if $-\gamma<1$. We consider the
model (10) with $0<\beta=-\gamma<1$. Let us change the integration
variable in eq.(45)
\begin{equation}
{\bf k}=\vert{\bf x}-{\bf x}^{\prime}\vert^{-1}\sqrt{r}{\bf q}
\end{equation}
Then, after an estimate of the remainder
\begin{equation} \exp(-W({\bf x}-{\bf
x}^{\prime}))\geq\exp(-cr{\bf p}^{2} \vert{\bf x}-{\bf
x}^{\prime}\vert^{2\beta})
\end{equation}
As a consequence \begin{displaymath}\begin{array}{l} \langle
\tilde{T}_{\tau}({\bf x},{\bf p})\tilde{T}_{\tau}({\bf
x}^{\prime},{\bf p}^{\prime})\rangle\cr \geq \delta({\bf p}+{\bf
p}^{\prime})\tilde{m}_{0}({\bf p}) (\mu^{2}{\bf p}^{2} +c{\bf
p}^{2} \vert{\bf x}-{\bf x}^{\prime}\vert^{2\beta})^{-1}
\Big(1-\exp(-\tau (\mu^{2}{\bf p}^{2} +c{\bf p}^{2} \vert{\bf
x}-{\bf x}^{\prime}\vert^{2\beta}))\Big)
\end{array}\end{displaymath}
Hence, for large $\vert{\bf x}-{\bf x}^{\prime}\vert$ we obtain
\begin{equation} \langle \tilde{T}_{\infty}({\bf
x},{\bf p})\tilde{T}_{\infty}({\bf x}^{\prime},{\bf
p}^{\prime})\rangle\geq \delta({\bf p}+{\bf
p}^{\prime})\tilde{m}_{0}({\bf p}) c^{-1}{\bf p}^{-2} \vert{\bf
x}-{\bf x}^{\prime}\vert^{-2\beta}
\end{equation}
This lower bound for the Kraichnan model is the same as the upper
bound (44) (here $\alpha=1$).

 Let
us calculate the expectation value in the exponential of eq.(38)
(denoted by $-W$) for the general $G$ of eq.(19)
\begin{equation}
\begin{array}{l}
W({\bf x}-{\bf x}^{\prime})=r^{2-\alpha-\gamma}\int d\omega d{\bf
k}{\bf p}\tilde{G}\left(\omega,{\bf k}\right){\bf p} \cr
\Big(2\left(\frac{1}{2}\mu^{2}{\bf
k}^{2}-i\omega\right)^{-1}\left(1-\left(\frac{1}{2}\mu^{2}{\bf
k}^{2}-i\omega\right)^{-1}\left(1-\exp\left(-\frac{1}{2}\mu^{2}{\bf
k}^{2}+i\omega\right)\right)\right)\cr
-\left(\frac{1}{4}\mu^{4}\vert{\bf
k}\vert^{4}+\omega^{2}\right)^{-1}\exp\left(i{\bf
k}r^{-\frac{1}{2}}\left({\bf x}- {\bf x}^{\prime}\right)\right)
\vert 1 -\exp\left(-\frac{1}{2}\mu^{2}{\bf
k}^{2}+i\omega\right)\vert^{2}\Big)
\end{array}
\end{equation}
We estimate this integral at ${\bf x}={\bf x}^{\prime}$ first.
Similarly as in eq.(46)  the scale invariance (21) leads to
\begin{equation} W({\bf 0})\geq cr^{2-\alpha-\gamma}{\bf p}^{2}
\end{equation}  if
\begin{displaymath}
\int d{\bf k}d\omega \tilde{G}\left(\omega,{\bf k}\right)
\left(\frac{1}{2}\mu^{2}{\bf k}^{2}-i\omega\right)^{-1}\theta
(\vert{\bf k}\vert -\frac{1}{\mu})<\infty
\end{displaymath} and
\begin{displaymath}
 \int\int d{\bf k}d\omega \tilde{G}\left(\omega,{\bf k}\right)
\left(\frac{1}{4}\mu^{4}\vert{\bf
k}\vert^{4}+\omega^{2}\right)^{\frac{1}{2}}\theta(\frac{1}{\mu}-\vert
{\bf k}\vert)<\infty \end{displaymath} Hence, under the assumption
(50) on the basis of the inequalities (52) and (58) we have the
lower bound ( generalizing that of eq.(51) to $\alpha\neq
1$)\begin{equation}\langle \tilde{T}_{\infty}({\bf x},{\bf
p})\tilde{T}_{\infty}({\bf x},{\bf p}^{\prime})\rangle\geq
\delta({\bf p}+{\bf p}^{\prime})\tilde{m}_{0}({\bf p})
\Big(c\theta(\vert {\bf p}\vert -\frac{1}{\mu})\vert{\bf
p}\vert^{-2+2\Omega}+c^{\prime}\theta(\frac{1}{\mu}-\vert {\bf
p}\vert )\vert {\bf
p}\vert^{-\frac{2-2\Omega}{2-\alpha-\gamma}}\Big)
\end{equation}
(at $\Omega=0$ this lower bound coincides with the upper bound
(40)).
 Next, if $\vert{\bf x}-{\bf
x}^{\prime}\vert$ is large then for $0<-\gamma=\beta<1$ we obtain
from eqs.(21) and (57) the lower bound
\begin{displaymath}
\exp(-W({\bf x}-{\bf x}^{\prime}))\geq\exp(-c{\bf p}^{2}r^{2-
\alpha }\vert{\bf x}-{\bf x}^{\prime}\vert^{2\beta})
\end{displaymath}
where the form of the rhs comes from a change of variables ${\bf
k}={\bf k}^{\prime}\vert{\bf x}-{\bf x}^{\prime}\vert^{-1}$ and
$\omega=\omega^{\prime} \vert {\bf x}-{\bf x}^{\prime}\vert^{-2}$
and an estimate of the remainder in eq.(57).

If we restrict ourselves to $G$ of the form (10) and $2\beta\geq
1$ then we can derive a more precise lower bound for $\exp(-W)$
with an application of the H\"older inequality
\begin{displaymath}
\vert {\bf x}+{\bf y}\vert^{2\beta}\leq 2^{2\beta-1} (\vert {\bf
x}\vert^{2\beta}+\vert {\bf y}\vert^{2\beta})
\end{displaymath}
From eq.(38) and the H\"older inequality we obtain after an
elementary calculation of the expectation value over the Brownian
paths
\begin{equation}
\begin{array}{l}
\exp(-W({\bf x}-{\bf x}^{\prime})) \geq\exp \Big( -C
r^{2-\alpha-\gamma} {\bf p}^{2} -cr^{2-\alpha}\vert {\bf x}-{\bf
x}^{\prime}\vert^{2\beta}{\bf p}^{2}\Big)
\end{array}
\end{equation}
Hence, after a calculation of the expectation value  in the
exponential in eq.(38) the remaining $r$ and the $a$ integrals
(from the representation (24)) in the correlation function (38)
read
\begin{displaymath}\begin{array}{l} \int_{0}^{\infty}dr\int da
\nu_{1}(a)\exp(-W -a\vert {\bf x}-{\bf
x}^{\prime}\vert^{2}-2\mu^{2}ra) \cr\geq
\frac{1}{2}\int_{0}^{\infty}dr\int da \nu_{1}(a)\exp(-W -a\vert
{\bf x}-{\bf x}^{\prime}\vert^{2})+
\frac{1}{2}\int_{0}^{\infty}dr\int da \nu_{1}(a)\exp(-W
-2\mu^{2}ra)
\end{array}\end{displaymath} where $\exp(-W) $ is lower bounded by
eq.(60). An easy estimate of this integral leads to the following
inequality  large ${\bf x}-{\bf x}^{\prime}$\begin{equation}
\begin{array}{l}\langle \tilde{T}_{\infty}({\bf x},{\bf
p})\tilde{T}_{\infty}({\bf x}^{\prime},{\bf
p}^{\prime})\rangle\cr\geq(\delta({\bf p}+{\bf
p}^{\prime})\tilde{m}_{0}({\bf p}) (K_{1}\vert{\bf
p}\vert^{-\frac{2-2\Omega}{2-\alpha}} \vert{\bf x}-{\bf
x}^{\prime}\vert^{-2\sigma}+K_{2}\vert{\bf
p}\vert^{-\frac{2}{2-\alpha}} \vert{\bf x}-{\bf
x}^{\prime}\vert^{-\frac{2\beta}{2-\alpha}-2\Omega})
\end{array}\end{equation} where
\begin{equation}
\sigma=\frac{\beta(1-\Omega)}{2-\alpha}
\end{equation}
For $0<2\beta<1$ this lower bound coincides with the upper bound
(44)(derived for $\Omega=0$). We expect that eq.(61) gives the
asymptotic behavior of the two-point correlation function for any
$0<2\beta<2$ because such a behavior is a consequence of a formal
exchange of the limit $\vert {\bf x}-{\bf
x}^{\prime}\vert\rightarrow \infty$ with the integral over $t$ and
the expectation value over the Brownian motion in eq.(34).

The lower bound (59) for small ${\bf p}$ is obtained by neglecting
the $\vert {\bf x}-{\bf x}^{\prime}\vert$-dependent term on the
rhs of eq.(60). We can see from eq.(59) that if
$\tilde{m}_{0}({\bf p})\simeq \vert {\bf p}\vert^{-\nu}$ and
$m_{1}({\bf k})\simeq \vert{\bf k}\vert^{-d +2\Omega}$ then the
$\langle \tilde{T}\tilde{T}\rangle$ correlations behave as $\vert
{\bf p}\vert^{-2-\nu+2\Omega}$ for large momenta (short distances
in the ${\bf z}$ direction), whereas the low momentum behaviour
(large  distance ) is $\vert {\bf
p}\vert^{-\nu-\frac{2-2\Omega}{2-\alpha-\gamma}}$. These estimates
show the effect of the random flow on the temperature correlations
 in the ${\bf z}$ direction. The effect on the temperature
correlations in the ${\bf x}$ direction is described by the lower
bound (61) and the upper bound (44). Again the decay of
temperature correlations is determined by scaling indices of the
velocity and source correlations.

\section{Higher order correlation functions}
Let us consider the multi-point correlation functions
\begin{equation}
\begin{array}{l}
\langle \tilde{T}_{\tau}({\bf x}_{1},{\bf
p}_{1})......\tilde{T}_{\tau}({\bf x}_{2n},{\bf
p}_{2n})\rangle=\sum_{pairs} \int_{0}^{\tau}dt_{1}....dt_{2n}\cr
\prod_{(j,k)}\delta({\bf p}_{j}+{\bf p}_{k})\delta(t_{j}-t_{k})
\exp(-\frac{1}{2}\mu^{2}\sum_{j}{\bf p}_{j}^{2}(\tau-t_{j}))\cr
E[\prod_{(j,k)}\tilde{m}({\bf x}_{j}-{\bf x}_{k}+\mu{\bf
b}_{j}(\tau-t_{j})-\mu{\bf b}_{k}(\tau-t_{k}),{\bf p}_{j})\cr
\exp\Big(-\frac{1}{2}\sum_{il}
\int_{0}^{\tau-t_{i}}\int_{0}^{\tau-t_{l}}ds ds^{\prime} {\bf
p}_{i} G_{0}(s-s^{\prime},{\bf x}_{i}-{\bf x}_{l}+\mu{\bf
b}_{i}(s)-\mu{\bf b}_{l}(s^{\prime})){\bf p}_{l}\Big)]
\end{array}
\end{equation}where the
sum is over all pairings in accordance with the Gaussian
combinatorics. From (63) we have
\begin{displaymath}
\begin{array}{l}
\vert\langle \tilde{T}_{\tau}({\bf x}_{1},{\bf
p}_{1})......\tilde{T}_{\tau}({\bf x}_{2n},{\bf
p}_{2n})\rangle\vert\leq\sum_{pairs}
\int_{0}^{\tau}dt_{1}....dt_{2n}\cr \prod_{(j,k)}\delta({\bf
p}_{j}+{\bf p}_{k})\delta(t_{j}-t_{k})
\exp(-\frac{1}{2}\mu^{2}\sum_{j}{\bf p}_{j}^{2}(\tau-t_{j}))\cr
E[\prod_{(j,k)}\vert\tilde{m}({\bf x}_{j}-{\bf x}_{k}+\mu{\bf
b}_{j}(\tau-t_{j})-\mu{\bf b}_{k}(\tau-t_{k}),{\bf
p}_{j})\vert]<\infty
\end{array}
\end{displaymath}
Hence, the equilibrium limit $\tau\rightarrow \infty$ exists.

If $m$ is either of the form (24) or (25) then we can apply the
Jensen inequality to the expectation value in the form $E[\exp
f]\geq \exp E[f]$. We obtain an analogue of the lower bound (38).
For the upper bound we apply the Jensen inequality to the time
integral\begin{equation}
\begin{array}{l}\exp(-\frac{1}{2}\int_{0}^{\tau}\int_{0}^{\tau}dsds^{\prime}
\int \int J_{k}(s) J_{l}(s^{\prime})\langle
v_{k}(s)v_{l}(s^{\prime})\rangle)\cr \leq \tau^{-2}
\int_{0}^{\tau}\int_{0}^{\tau}dsds^{\prime}\exp(-\frac{\tau^{2}}{2}
\int \int J_{k}(s) J_{l}(s^{\prime})\langle
v_{k}(s)v_{l}(s^{\prime})\rangle)\end{array}
\end{equation}
where
\begin{displaymath} {\bf J}(s,{\bf
u})=-\theta(s)\sum_{k=1}^{2n}{\bf p}_{k}\delta({\bf u}-{\bf
x}_{k}-\mu{\bf b}_{k}(\tau-s))\end{displaymath} and the additional
integral in eq.(64) is over the spatial variable ${\bf u}$.

We can repeat the basic estimates concerning the behavior for low
${\bf z}$ momenta and large ${\bf x}$ distances by means of the
methods applied for the two-point correlations. First, by means of
the Jensen inequalities we reduce the estimates of the expectation
values to finite dimensional integrals. From the Jensen
inequalities we can see that the correlation functions are bounded
in $\tau$ when $\tau\rightarrow\infty$. Next, the results
concerning the scaling behavior for $2n$-point functions can be
obtained  by an introduction of spherical coordinates in the
$dt_{1}...dt_{n}$ integral in eq.(63). Then, the correlation
functions scale in a simple way with respect to the temporal
radius $r$. Let us explain such estimates in more detail for
$n=2$. Then,
\begin{equation}
\begin{array}{l}
\langle \tilde{T}_{\tau}({\bf x}_{1},{\bf
p}_{1})......\tilde{T}_{\tau}({\bf x}_{4},{\bf
p}_{4})\rangle=\delta({\bf p}_{1}+{\bf p}_{3}) \delta({\bf
p}_{2}+{\bf p}_{4}) \cr\int_{0}^{\tau}dt_{1}\int_{0}^{\tau}dt_{2}
\exp(-\mu^{2}{\bf p}_{1}^{2}(\tau-t_{1})-\mu^{2}{\bf
p}_{2}^{2}(\tau-t_{2}))\cr E[\tilde{m}({\bf x}_{1}-{\bf
x}_{3}+\mu{\bf b}_{1}(\tau-t_{1})-\mu{\bf b}_{3}(\tau-t_{1}),{\bf
p}_{1})\cr \tilde{m}({\bf x}_{2}-{\bf x}_{4}+\mu{\bf
b}_{2}(\tau-t_{2})-\mu{\bf b}_{4}(\tau-t_{2}),{\bf p}_{2})\cr
\exp\Big(-\sum_{j=1,2}
\int_{0}^{\tau-t_{j}}\int_{0}^{\tau-t_{j}}dt dt^{\prime} {\bf
p}_{j} G_{0}(t-t^{\prime},\mu{\bf b}_{j}(t)-\mu{\bf
b}_{j}(t^{\prime})){\bf p}_{j}\cr +\sum_{j< k}
\int_{0}^{\tau-t_{j}}\int_{0}^{\tau-t_{k}}dt dt^{\prime} {\bf
p}_{j} G_{0}(t-t^{\prime},{\bf x}_{j}-{\bf x}_{k}+\mu{\bf
b}_{j}(t)-\mu{\bf b}_{k}(t^{\prime})){\bf p}_{k}\Big)] + permut.
\end{array}
\end{equation}
where the sum is over permutations of the numbers from 1 to 4 in
accordance with the Gaussian combinatorics; in the sum in the
exponential we set $t_{1}=t_{3}$ and $t_{2}=t_{4}$. Let
$\tau-t_{1}=r\cos\theta$ , $\tau-t_{2}=r\sin\theta$,
$t=r\sigma\cos\theta$ and $t^{\prime}=r\sigma^{\prime}\sin\theta$.
In such a case $r$ scales in the exponential in the same way as in
eqs.(37)-(38). The integral $dt_{1}dt_{2}=drr d\theta$ adds an
additional power of r. Under the assumption (31) the small ${\bf
p}$ behavior of the correlation functions (65) at $ \tau=\infty$
is determined by the integral
\begin{equation}
\begin{array}{l}
\vert\langle \tilde{T}_{\infty}({\bf x}_{1},{\bf
p}_{1})......\tilde{T}_{\infty}({\bf x}_{4},{\bf
p}_{4})\rangle\vert \cr \simeq \vert\tilde{m}_{0}\vert({\bf
p}_{1})\vert\tilde{m}_{0}\vert({\bf
p}_{2})\int_{0}^{\infty}drrE[\exp(-r^{2-\alpha-\gamma}\sum_{jk}{\bf
p}_{j}G_{0}{\bf p}_{k} )]+permut.
\cr\simeq\vert\tilde{m}_{0}\vert({\bf
p}_{1})\vert\tilde{m}_{0}\vert({\bf p}_{2})E[\vert\sum_{jk}{\bf
p}_{j}G_{0}{\bf p}_{k}\vert^{-\frac{2}{2-\alpha-\gamma}}]+permut.
\end{array}
\end{equation}
For large distances , $\gamma=-\beta<0$ and  $G$ of eq.(10) we can
expand the dependence on the Brownian motion in eq.(66) in powers
of $\mu\vert {\bf x}_{j}-{\bf x}_{k}\vert^{-1}$. The leading order
reads
\begin{equation}
\begin{array}{l}
\vert\langle \tilde{T}_{\infty}({\bf x}_{1},{\bf
p}_{1})......\tilde{T}_{\infty}({\bf x}_{4},{\bf
p}_{4})\rangle\vert \cr\simeq\vert\tilde{m}_{0}\vert({\bf
p}_{1})\vert\tilde{m}_{0}\vert({\bf p}_{2})\vert\sum_{jk}{\bf
p}_{j}{\bf p}_{k}\vert {\bf x}_{j}-{\bf
x}_{k}\vert^{2\beta}\vert^{-\frac{2}{2-\alpha}}+permut.
\end{array}
\end{equation}
 Note that
the power describing the low ${\bf p}$ behavior in eq.(66) and
large ${\bf x}$ behavior in eq.(67) is twice as big as that for
the two-point function (34) and (40) indicating the asymptotic
scale invariance of the temperature $\tilde{T}_{\infty}({\bf
x},{\bf p})$ at low momenta or large distances. This property can
be extended to the $2n$ correlation functions where the scaling
index is proportional to $n$ as a consequence of the $drr^{n-1}$
time integral in the spherical time coordinates. Such a behavior
of the integrals suggests that if the velocities and the sources
are scale invariant then the temperatures scale at large distances
with the scale dimension determined by the two-point function.
\section{Discussion}
The power-law behavior of turbulent velocity correlation functions
and passive scalar correlation functions in a homogeneous
isotropic turbulent flow has been widely discussed in the
literature since the basic papers of Kolmogorov \cite{kolmogorov}
followed by Obukhov \cite{obukhov},Corrsin\cite{corsin} and
Batchelor\cite{batchelor} (concerning the scalar advection). The
universal Kolmogorov $\frac{5}{3}$ law for spectral velocity
distribution as well as passive scalar distribution is derived by
means of dimensional arguments (independent of any dynamical
model). A  statistical homogeneity and isotropy of the turbulence
at a microscale in a sufficiently large space interval (called the
" inertial range") is at the base of the Kolmogorov theory. Under
these assumptions the velocity (or passive scalar) correlation
functions are universal ,i.e.,independent of the source
distribution $m$. An experimental verification is not simple.
Turbulent flows are usually non-homogeneous and non-isotropic at a
macro scale. However, if a flow satisfying Kolmogorov assumptions
is created then the spectral Kolmogorov law is satisfied  in the
inertial range \cite{goto}. Nevertheless, it is common for flows
in nature that Kolmogorov assumptions are not satisfied (for some
studies of such turbulent flows see
\cite{celani}\cite{mydlarski}). Even if the velocity is satisfying
the Kolmogorov law the analogous Obukhov law for $\rho$ may fail
\cite{celani}\cite{mydlarski}\cite{anton}. As the authors in
\cite{celani} point out some problems with the verification of
Kolmogorov's theory concerns a construction of a flow which would
be homogeneous and isotropic in a sufficiently large inertial
range ( usually boundary conditions or sources violate a global
symmetry). They suggest a study of non-isotropic flows.

An investigation of a general class of dynamical models of
randomly forced Navier-Stokes and passive scalar equations is
still beyond the reach of analytical as well as numerical methods.
A substantial progress has been achieved in the white noise
randomly forced passive scalar (Kraichnan model)\cite{kraichnan}
\cite{gawedzki}\cite{falk}. However, the white noise distribution
of velocities is quite unrealistic. Our main motivation in these
studies was a derivation of the scaling behavior for velocities
which are not of the white noise type. A passive scalar in a shear
flow independent of the coordinates in the direction of the flow
was studied before in \cite{majda}\cite{glimm}. However, these
authors were interested in the anomalous free decay of solutions
of the advection-diffusion equation.

 Our results  predict a power-law of the passive scalar correlations in
non-isotropic flows. The results depend on the source distribution
$m$ because the source $f$ is present at any scale. We do not
specify any inertial range in our model. In general, the
correlations must depend on the source (for a discussion of random
forcing see \cite{stephen}). This can be seen from the detailed
calculations in \cite{gawedzki2}\cite{proc} performed in the
isotropic Kraichnan model (white noise in
time)\cite{kraichnan}\cite{gawedzki}\cite{falk}. The two-point
passive scalar  correlations depend explicitly on the source and
on the molecular diffusivity $\mu^{2}$. Only in a proper limit of
the source covariance $m$ and $\mu\rightarrow 0$ the universal
scaling law comes out.

Before we  summarize our results let us begin with  simple models.
First, consider a pure diffusion corresponding to ${\bf V}=0$.
Then
\begin{displaymath}
\begin{array}{l}
\langle \tilde{T}_{\tau}({\bf x},{\bf p})\tilde{T}_{\tau}({\bf
x}^{\prime},{\bf p}^{\prime})\rangle= 2\delta({\bf p}+{\bf
p}^{\prime})\int_{0}^{\tau}dr\exp(-\mu^{2}{\bf p}^{2}r) \cr
E[\tilde{m}( {\bf x}-{\bf x}^{\prime}+\mu\sqrt{r}{\bf
b}(1)-\mu\sqrt{r}{\bf b}^{\prime}(1),{\bf p}) ]\cr = 2\delta({\bf
p}+{\bf
p}^{\prime})(2\pi)^{-D+d}\int_{0}^{\tau}dr\exp(-\mu^{2}{\bf
p}^{2}r) \cr \int d{\bf u}d{\bf w}\exp(-\frac{{\bf
u}^{2}}{2}-\frac{{\bf w}^{2}}{2})\tilde{m}( {\bf x}-{\bf
x}^{\prime}+\mu\sqrt{r}{\bf u}-\mu\sqrt{r}{\bf w},{\bf p}) ]\cr
=\mu^{-2}\delta({\bf p}+{\bf p}^{\prime})\int d{\bf k} \exp(i{\bf
k}({\bf x}-{\bf x}^{\prime}))\tilde{m}_{1}({\bf
k})\tilde{m}_{0}({\bf p})({\bf p}^{2}+{\bf
k}^{2})^{-1}\cr\Big(1-\exp(-\mu^{2}({\bf p}^{2}+{\bf
k}^{2})\tau)\Big)\end{array}
\end{displaymath}
 In the limit
$\tau\rightarrow \infty$
\begin{equation}
\langle T_{\infty}({\bf x},{\bf z})T_{\infty}({\bf
x}^{\prime},{\bf z}^{\prime})\rangle= \mu^{-2}\int d{\bf k} d{\bf
p}\exp(i{\bf k}({\bf x}-{\bf x}^{\prime}))\exp(i{\bf p}({\bf
z}-{\bf z}^{\prime}))\tilde{m}({\bf k},{\bf p})({\bf p}^{2}+{\bf
k}^{2})^{-1} \end{equation} Hence\begin{equation}
\rho_{\infty}({\bf k},{\bf p})=\mu^{-2}({\bf p}^{2}+{\bf
k}^{2})^{-1}\tilde{m}_{1}({\bf k})\tilde{m}_{0}({\bf p})
\end{equation}
  Let us note that the behavior of the temperature
correlations changes abruptly for large $\vert{\bf z}-{\bf
z}^{\prime}\vert$
 at $\tau=\infty$  in this simple model. At finite
$\tau$ it is the same as that of the source (say $\vert{\bf
z}-{\bf z}^{\prime}\vert^{-d+\nu}$) whereas at $\tau=\infty$ it
becomes $\vert{\bf z}-{\bf z}^{\prime}\vert^{-d+\nu+2}$. However,
it can be seen from eq.(68) that after the limit $\tau\rightarrow
\infty$ the limit $\mu\rightarrow 0$ does not exist in the model
without the advection. If we first take $\mu\rightarrow 0$ then
the subsequent limit $\tau\rightarrow \infty$ is linearly
divergent in $\tau$. The strong $\mu$-dependence of the asymptotic
behavior means that this parameter sets a scale on time and space
which determines different scaling behavior. In Appendix A we show
that  the limits $\mu\rightarrow 0$ and $\tau\rightarrow \infty$
can be interchanged in the model with a random advection . The
correlation functions ${\cal S}^{(2n)}$ in a non-isotropic
Kraichnan model are discussed in Appendix B. The correlation
functions ${\cal S}^{(2n)}({\bf x}_{1},{\bf p}_{1},.....,{\bf
x}_{2n},{\bf p}_{2n})$ can be calculated exactly in the limit
$\mu\rightarrow 0$ (eq.(86)). They show no anomalous scaling
(encountered in the isotropic model
\cite{kraichnan}\cite{gawedzki}) as long as the points ${\bf
x}_{j}$ are different. The scaling behavior can change after a
transformation to the configuration space (the Fourier transform
does not exist in the usual sense).

 Let us  compare the two-point temperature correlation function
 (68)  with the one
 in a random flow which is bounded in space and time
  ,i.e., $G=\langle {\bf v}{\bf v}\rangle\simeq const$.
Under the assumption (31) we obtain

\begin{equation}
\begin{array}{l}
\langle \tilde{T}_{\infty}({\bf x},{\bf p})\tilde{T}_{\infty}({\bf
x}^{\prime},{\bf p}^{\prime})\rangle\simeq K\delta({\bf p}+{\bf
p}^{\prime})\tilde{m}_{0}({\bf
p})\int_{0}^{\infty}dr\exp(-\mu^{2}{\bf p}^{2}r -c{\bf p}^{2}r^{2}
)\end{array}
\end{equation}
 The integral (70) behaves as $\tilde{m}_{0}({\bf p}){\bf
p}^{-2}$ for large ${\bf p}$ and as $\tilde{m}_{0}({\bf
p})\vert{\bf p}\vert^{-1}$ for a small ${\bf p}$ in agreement with
eq.(59) for $\Omega=\alpha=\gamma=0$. Our results of secs.4 and 5
give an extension of the simple observations on the temperature
correlation functions derived in this section for a pure diffusion
and for an advection by a uniformly bounded random flow.

In our model (defined by the assumption that the velocity does not
depend on coordinates in the direction of the flow)  the spectral
distribution in the corresponding momentum is proportional to the
source distribution $\tilde{m}$ as can be seen from eq.(34). We
could consider a source $f$ with the covariance $m({\bf x},{\bf
z})$ which (approximately in a certain range as in
refs.\cite{gawedzki2}\cite{proc}) is independent of ${\bf x}$. In
such a case the spectral equilibrium distribution (6) for a pure
diffusion $\rho_{\infty}({\bf k},{\bf p})$ (67) is $\delta({\bf
k})\tilde{m}_{0}({\bf p}){\bf p}^{-2}$ where the ${\bf p}^{-2}$
behavior comes from the molecular diffusivity. The temperature
correlations remain independent of ${\bf x}$ and the limit
$\mu\rightarrow 0$ does not exist. A random advection is changing
the behavior of temperature correlations in ${\bf x}$ as well as
in ${\bf p}$. This change involves a non-perturbative mechanism
which could not be seen in an expansion in ${\bf V}$. It comes
from an exponential of $G$ in eq.(34). In particular,  a steady
flow bounded in ${\bf x}$ gives $\rho_{\infty}({\bf k},{\bf
p})=\delta({\bf k})\vert {\bf p}\vert^{-1}$ for $\vert {\bf
p}\vert\ll \frac{1}{\mu}$ whereas for the random velocity growing
in space with the index $\beta$ (eq.(10)) we have for a  small
${\bf k}$
 the behavior $\rho_{\infty}({\bf k},{\bf p})\simeq \vert {\bf
k}\vert^{-d+\frac{2\beta}{2-\alpha}} $ as follows from eq.(61).

In experiments ($D=3$) we could create an anisotropic  flow with
the Kolmogorov index (10) $\beta=\frac{1}{3}$ in $d=2$ or $d=1$.
In such a case we obtain definite predictions concerning the
temperature distribution. This will be \newline $ \vert {\bf
x}-{\bf x}^{\prime}\vert^{-\frac{2}{3}}\vert{\bf z}-{\bf
z}^{\prime}\vert^{\nu}$ where
\begin{displaymath}
\nu= \frac{2}{2-\alpha}-(D-d)
\end{displaymath}
and $D-d$  is either $1$ or $2$ and there is a restriction $\alpha
-\beta<1$ coming from the requirement of the integrability of the
expression in the exponential of (34).

In general, we can see from eqs.(40),(42),(44),(59) and (61) that
the turbulent behavior $\gamma=-\beta<0$ of the velocity field
will (in comparison to pure diffusion) decrease the temperature
correlations in the direction orthogonal to the flow and increase
the correlations (at the fixed $\alpha$)in the direction of the
flow. These effects contribute to the more coherent heat
distribution in a turbulent stream.
\section*{Appendix A: The limit
$\mu\rightarrow 0$} If there is no diffusion ($\mu=0$) then  our
formulas in secs.4-6 at finite $\tau$ remain valid but need some
interpretation. There is no expectation value over the Brownian
motion. In such a case in some formulas (as in eqs.(34)-(35))
$\gamma=0$. Let us consider as an example the formula (34) at
$\mu=0$
\begin{equation}
\begin{array}{l}
\langle \tilde{T}_{\tau}({\bf x},{\bf p})\tilde{T}_{\tau}({\bf
x}^{\prime},{\bf p}^{\prime})\rangle=  \delta({\bf p}+{\bf
p}^{\prime})\tilde{m}({\bf x}-{\bf x}^{\prime},{\bf
p})\int_{0}^{\tau}dt  \exp(-i{\bf p}\int_{0}^{\tau-t}ds{\bf
U}(\tau -s,{\bf x}))\cr \exp\Big(-(\tau-t)^{2-\alpha}
\int_{0}^{1}\int_{0}^{1}d\sigma d\sigma^{\prime} {\bf p}
G_{0}(\sigma-\sigma^{\prime},{\bf 0}){\bf p} \cr
+(\tau-t)^{2-\alpha} \int_{0}^{1}\int_{0}^{1}d\sigma
d\sigma^{\prime} {\bf p} G_{0}(\sigma-\sigma^{\prime},{\bf x}-{\bf
x}^{\prime}){\bf p}\Big)
\end{array}
\end{equation}
For the Kraichnan model \cite{kraichnan} (35) the formula (71)
reads (with the Stratonovitch interpretation of the gradient term,
see the discussion at the beginning of sec.2).

\begin{equation}
\begin{array}{l}
\langle \tilde{T}_{\tau}({\bf x},{\bf p})\tilde{T}_{\tau}({\bf
x}^{\prime},{\bf p}^{\prime})\rangle= \delta({\bf p}+{\bf
p}^{\prime})\tilde{m}({\bf x}-{\bf x}^{\prime},{\bf
p})\int_{0}^{\tau}dt  \exp(-i{\bf p}\int_{0}^{\tau-t}{\bf U}(\tau
-s,{\bf x})ds)\cr \exp\Big(-(\tau-t) {\bf p} D_{0}({\bf 0}){\bf p}
+(\tau-t) {\bf p} D_{0}({\bf x}-{\bf x}^{\prime}){\bf p}\Big)
\end{array}
\end{equation}
In the limit $\tau \rightarrow \infty$ and for ${\bf U}=0$ we can
calculate the integral over time in eq.(71) with the result
\begin{equation}
\begin{array}{l}
\langle \tilde{T}_{\infty}({\bf x},{\bf p})\tilde{T}_{\infty}({\bf
x}^{\prime},{\bf p}^{\prime})\rangle= C\delta({\bf p}+{\bf
p}^{\prime})\tilde{m}({\bf x}-{\bf x}^{\prime},{\bf p})\cr\Big(
\int_{0}^{1}\int_{0}^{1}d\sigma d\sigma^{\prime} {\bf p}
G_{0}(\sigma-\sigma^{\prime},{\bf 0}){\bf p} -
\int_{0}^{1}\int_{0}^{1}d\sigma d\sigma^{\prime} {\bf p}
G_{0}(\sigma-\sigma^{\prime},{\bf x}-{\bf x}^{\prime}){\bf
p}\Big)^{-\frac{1}{2-\alpha}}
\end{array}
\end{equation}
in agreement with the bounds (56) and (61). We can also calculate
the higher order correlation functions. As an example, the four
point function (65) reads
\begin{displaymath}
\begin{array}{l}
\langle \tilde{T}_{\tau}({\bf x}_{1},{\bf
p}_{1})......\tilde{T}_{\tau}({\bf x}_{4},{\bf p}_{4})\rangle\cr =
 \delta({\bf p}_{2}+{\bf p}_{4})  \delta({\bf
p}_{1}+{\bf p}_{3})\tilde{m}({\bf x}_{1}-{\bf x}_{3},{\bf
p}_{1})\tilde{m}({\bf x}_{2}-{\bf x}_{4},{\bf
p}_{2})\cr\int_{0}^{\tau}dt_{1}\int_{0}^{\tau}dt_{2}
\exp\Big(-\sum_{j=1,2}
\int_{0}^{\tau-t_{j}}\int_{0}^{\tau-t_{j}}dt dt^{\prime} {\bf
p}_{j} G_{0}(t-t^{\prime},{\bf 0}){\bf p}_{j}\cr +\sum_{j< k}
\int_{0}^{\tau-t_{j}}\int_{0}^{\tau-t_{k}}dt dt^{\prime} {\bf
p}_{j} G_{0}(t-t^{\prime},{\bf x}_{j}-{\bf x}_{k}){\bf p}_{k}\Big)
+ permut.
\end{array}
\end{displaymath}
 We can obtain detailed estimates of the
time integrals for any $\alpha$. In some special cases the
integrals can be explicitly calculated. In Appendix B we give the
formula (eq.(88)) for the Kraichnan model ($\alpha=1$). For a
steady flow ($\Gamma(s)=1$ in eq.(10),$\alpha=0$) at $\tau=\infty$
the integration over $t_{j}$ gives
\begin{equation}
\begin{array}{l}
\langle \tilde{T}_{\infty}({\bf x}_{1},{\bf
p}_{1})......\tilde{T}_{\infty}({\bf x}_{4},{\bf p}_{4})\rangle\cr
=
 \delta({\bf p}_{2}+{\bf p}_{4})  \delta({\bf
p}_{1}+{\bf p}_{3})\tilde{m}({\bf x}_{1}-{\bf x}_{3},{\bf
p}_{1})\tilde{m}({\bf x}_{2}-{\bf x}_{4},{\bf p}_{2}) \cr
\Big(4{\bf p}_{1}^{2}{\bf p}_{2}^{2}D_{0}({\bf x}_{1}-{\bf
x}_{3})D_{0}({\bf x}_{2}-{\bf x}_{4})\cr - ({\bf p}_{1}{\bf
p}_{2})^{2}(D_{0}({\bf x}_{1}-{\bf x}_{4})+D_{0}({\bf x}_{2}-{\bf
x}_{3})+D_{0}({\bf x}_{1}-{\bf x}_{2}) +D_{0}({\bf x}_{3}-{\bf
x}_{4}))^{2}\Big)^{-\frac{1}{2}}+permut.
\end{array}
\end{equation}
where $D_{0}({\bf x}_{j}-{\bf x}_{k})=-\vert {\bf x}_{j}-{\bf
x}_{k}\vert^{2\beta}$ in the model (10).
\section*{Appendix B:The Kraichnan model}
If $\Gamma(t-t^{\prime})=\delta(t-t^{\prime})$ then we obtain a
closed set of equations for the correlation functions
\begin{equation}
{\cal S}_{\tau}^{(n)}({\bf x}_{1},...,{\bf x}_{n};{\bf
p}_{1},....,{\bf p}_{n})=\langle \tilde{T}_{\tau}({\bf x}_{1},{\bf
p}_{1})....\tilde{T}_{\tau}({\bf x}_{n},{\bf p}_{n})\rangle
\end{equation}
These equations have been derived by Kraichnan \cite{kraichnan}
for velocities depending on all coordinates. In our simplified
model (9)-(10) they read (the odd order correlation functions are
zero)
\begin{equation}
\begin{array}{l}
\partial_{\tau}{\cal S}_{\tau}^{(2n)}=
\frac{1}{2}\mu^{2}\sum_{j=1}^{j=2n}\triangle_{j}{\cal
S}_{\tau}^{(2n)}-\frac{1}{2}(\mu^{2}+D_{0}({\bf
0}))\sum_{j=1}^{j=2n}{\bf p}_{j}^{2}{\cal S}_{\tau}^{(2n)}\cr
+\sum_{<j,k>}{\bf p}_{j}D_{0}({\bf x}_{j}-{\bf x}_{k}){\bf
p}_{k}{\cal S}_{\tau}^{(2n)}+\sum_{<j,k>}\delta({\bf p}_{j}+{\bf
p}_{k})\tilde{m}({\bf x}_{j}-{\bf x}_{k},{\bf p}_{j}){\cal
S}_{\tau}^{(2n-2)}(jk)\cr \equiv {\cal M}S_{\tau}^{(2n)}+ {\cal
R}{\cal S}_{\tau}^{(2n-2)}
\end{array}
\end{equation}
 where $D_{0}$ is the translation invariant part of $D$
 and ${\cal S}(jk)$ means that the coordinates ${\bf x}_{j}$
and ${\bf x}_{k}$ are lacking in ${\cal S}$. The term $D({\bf 0})$
(adding to $\mu^{2}$) comes from the Stratonovitch interpretation
of eq.(1).The solution of eq.(76) reads
\begin{equation}
\begin{array}{l} {\cal S}_{\tau}^{(2n)} =\exp(\tau {\cal
M})S_{0}^{(2n)}+ \int_{0}^{\tau}dt\exp((\tau-t) {\cal M}){\cal
R}{\cal S}_{t}^{(2n-2)} \end{array} \end{equation} If the operator
${\cal M}$ is strictly negative in the space $L^{2}(R^{2dn})$ then
the limit $\tau\rightarrow \infty$ exists and does not depend on
the initial condition ${\cal S}_{0}^{(2n)}$.

We can express the solution of eq.(76) by means of the Feynman-Kac
formula for the heat kernel
\begin{equation}
(\exp (r{\cal M})g)({\bf x}_{1},.....,{\bf x}_{2n})=
E[\exp(\int_{0}^{r}ds W({\bf b}(s)))g({\bf x}_{1}+\mu{\bf
b}_{1}(r),...,{\bf x}_{2n}+\mu{\bf b}_{2n}(r))]
\end{equation}
where
\begin{equation}
\begin{array}{l}
W(s)=-\frac{1}{2}(\mu^{2}+D_{0}({\bf 0}))\sum_{j=1}^{j=2n}{\bf
p}_{j}^{2} +\sum_{<j,k>}{\bf p}_{j}D_{0}({\bf x}_{j}+\mu{\bf
b}_{j}(s)-{\bf x}_{k}-\mu{\bf b}_{k}(s)){\bf p}_{k} \end{array}
\end{equation} We obtain an upper bound on the correlation
functions (78) from the Jensen inequality as applied to the time
integral

\begin{equation}(\exp r{\cal
M})g)({\bf x}_{1},.....,{\bf x}_{2n})\leq
\frac{1}{r}\int_{0}^{r}ds E[\exp(r W({\bf b}(s)))\vert g\vert({\bf
x}_{1}+\mu{\bf b}_{1}(r),...,{\bf x}_{2n}+\mu{\bf b}_{2n}(r))]
\end{equation}
If $g=\exp h$ (or  a superposition with positive coefficients of
such functions as in eq.(25))then we have the lower bound from the
Jensen inequality as applied to the expectation value
\begin{equation}
(\exp (r{\cal M})\exp h)({\bf x}_{1},.....,{\bf x}_{2n})\geq \exp
E[\int_{0}^{r}ds W({\bf b}(s)))+h({\bf x}_{1}+\mu{\bf
b}_{1}(r),...,{\bf x}_{2n}+\mu{\bf b}_{2n}(r))]
\end{equation}
As an example, the formula for the two point function (in the
limit $\tau\rightarrow \infty$) with the velocity correlations
defined by eq.(10) reads
\begin{equation}
\begin{array}{l}{\cal S}_{\infty}^{(2)}({\bf x}_{1},{\bf x}_{2},{\bf p}_{1},{\bf
p}_{2})= \delta({\bf p}_{1}+{\bf
p}_{2})\int_{0}^{\infty}dr\exp(-r\mu^{2}{\bf p}_{1}^{2}) \cr
E[\exp(-{\bf p}_{1}^{2}\int_{0}^{r}ds\vert{\bf x}_{1}-{\bf
x}_{2}+\mu{\bf b}_{1}(s)-\mu{\bf b}_{2}(s)\vert^{2\beta})\tilde{
m}({\bf x}_{1}+\mu{\bf b}_{1}(r)-{\bf x}_{2}-\mu{\bf
b}_{2}(r),{\bf p}_{1})]
\end{array}
\end{equation}
 Then, the resulting correlation functions
are controlled from below and from above by the Jensen
inequalities. For the lower bound (81) we obtain an explicit
formula (using the representation (24) for $m_{1}$)
\begin{equation}
\begin{array}{l}{\cal S}_{\infty}^{(2)}({\bf x}_{1},{\bf x}_{2},{\bf p}_{1},{\bf
p}_{2})\geq \delta({\bf p}_{1}+{\bf p}_{2})\tilde{m}_{0}({\bf
p})\int d\nu_{1}(a)\int_{0}^{\infty}dr\exp(-r\mu^{2}{\bf
p}_{1}^{2}) \cr
 \exp(-{\bf p}^{2}r^{\beta +1}h(r^{-\frac{1}{2}}\vert {\bf x}_{1}-{\bf
 x}_{2}\vert)-a\vert{\bf x}_{1}-{\bf x}_{2}\vert^{2}-2\mu^{2}ra)
 \end{array}
 \end{equation}
where
\begin{equation}
\begin{array}{l}
h(\rho)=K\rho^{2(1+\beta)}\int_{0}^{\rho^{-2}}d\lambda\int_{0}^{\infty}
db  b^{-1-\beta}
 \Big(1-(1+2\mu^{2}\lambda
 b)^{-\frac{d}{2}}\exp(-\frac{b}{2(1+2\mu^{2}b\lambda)})\Big)
 \end{array}\end{equation}
 here $K$ is a positive constant.
From eq.(84) it can easily be seen that for large ${\bf x} -{\bf
x}^{\prime}$ (small $\rho$ in eq.(84)) the $r$-integrand in
eq.(83) behaves as
\begin{displaymath}
\exp(-K r{\bf p}^{2}\vert {\bf x}_{1}-{\bf
x}_{2}\vert^{2\beta}-a\vert {\bf x}_{1}-{\bf
x}_{2}\vert^{2}-2\mu^{2}ra)
\end{displaymath}
(as shown in another way in eq.(60);here $\alpha=1$)leading as a
consequence to the estimate (61) for the correlation functions. We
can continue the Jensen inequalities for higher correlation
functions as from eqs.(77) and (78) it follows that the
correlation functions are again in the form of superpositions of
exponentials.

For lower order correlations a direct study of the differential
equation (76) can be equally efficient. As an example, if $D=3$
and $d=2$ then the equation (76) at $\tau=\infty$ (with the
velocity covariance (10)) reads (here $\rho=\vert{\bf x}_{1}-{\bf
x}_{2}\vert$)

\begin{equation}
(\mu^{2}\frac{1}{\rho}\partial_{\rho}\rho\partial_{\rho}-\mu^{2}p^{2}-p^{2}\rho^{2\beta})
{\cal T}_{\infty}^{(2)}(\rho,p;\mu)=\tilde{m}(\rho,p)
\end{equation}
where we defined
\begin{displaymath}
{\cal S}^{(2)}(\rho,p_{1},p_{2};\mu)=\delta( p_{1}+p_{2}){\cal
T}^{(2)}(\rho,p_{1};\mu)
\end{displaymath}
In contradistinction to the spherically symmetric case
\cite{kraichnan} eq.(85) is not explicitly soluble but its
asymptotic solution (61) is easy to obtain. This asymptotic
behavior is the same as the limit $\mu=0$ of the solution (85)
\begin{equation}
{\cal T}_{\infty}^{(2)}(\rho,p;0)=-p^{-2}\rho^{-2\beta}
\tilde{m}(\rho,p)
\end{equation}
In general, from eq.(76) the limit $\mu=0$ can be obtained
inductively  \begin{equation}\begin{array}{l}{\cal
S}_{\infty}^{(2n)}({\bf x}_{1},...,{\bf x}_{2n};{\bf
p}_{1},...,{\bf p}_{2n};0)=\Big(\frac{1}{2}D_{0}({\bf
0})\sum_{j=1}^{j=2n}{\bf p}_{j}^{2}- \sum_{<j,k>}{\bf
p}_{j}D_{0}({\bf x}_{j}-{\bf x}_{k}){\bf
p}_{k}\Big)^{-1}\cr\sum_{<i,l>}\delta({\bf p}_{i}+{\bf
p}_{l})\tilde{m}({\bf x}_{i}-{\bf x}_{l},{\bf p}_{i}){\cal
S}_{\infty}^{(2n-2)}(il;0) \end{array}\end{equation} The formulas
for the asymptotic behavior (61) ($\alpha=1$) and (66)
($\alpha=1,\gamma=0$) agree with the exact solution (87). For
$n=1$ the solution (82) takes the form (86) whereas for $n=2$ we
have
\begin{equation}\begin{array}{l}{\cal
S}_{\infty}^{(4)}({\bf x}_{1},...,{\bf x}_{4};{\bf p}_{1},...,{\bf
p}_{4};0)=\Big(\frac{1}{2}D_{0}({\bf 0}))\sum_{j=1}^{j=4}{\bf
p}_{j}^{2}- \sum_{<j,k>}{\bf p}_{j}D_{0}({\bf x}_{j}-{\bf
x}_{k}){\bf p}_{k}\Big)^{-1}\cr\Big(\delta({\bf p}_{1}+{\bf
p}_{2})\delta({\bf p}_{3}+{\bf p}_{4})\big({\bf p}_{1}D_{0}({\bf
0}){\bf p}_{1}-{\bf p}_{1}D_{0}({\bf x}_{1}-{\bf x}_{2}){\bf
p}_{1}\big)^{-1}\cr\tilde{m}({\bf x}_{1}-{\bf x}_{2},{\bf
p}_{1})\tilde{m}({\bf x}_{3}-{\bf x}_{4},{\bf p}_{3})
+permut.\Big)
\end{array}\end{equation}
For the scale invariant random velocity field (10) $D_{0}({\bf
0})=0$ and $D_{0}({\bf x}_{j}-{\bf x}_{k})=-\vert{\bf x}_{j}-{\bf
x}_{k}\vert^{2\beta}$. It follows from eqs.(87)-(88) that the
temperature correlation functions are scale invariant under scale
transformations of the coordinates ${\bf x}_{j}$ as well as ${\bf
p}_{j}$. When $\mu=0$ then the correlation functions ${\cal
S}^{(2n)}_{\infty}$ are singular at coinciding points (the limit
$\mu\rightarrow 0$ has been studied earlier by other methods in
\cite{wei}\cite{hula}). The bound (40) is valid for $\mu>0$.

\end{document}